\documentclass[letterpaper,twocolumn,10pt]{article}
\usepackage{usenix,epsfig,endnotes}

\usepackage[misc]{ifsym} 
\usepackage{lipsum}
\usepackage{verbatim}  
\usepackage{comment}
\usepackage{algorithm}
\usepackage{algpseudocode}  
\usepackage{subfigure}
\usepackage{multirow}
\usepackage{booktabs}
\usepackage{graphicx}
\usepackage{threeparttable}
\usepackage{listings}

\begin{document}

\title{\Large \bf {Gamify Stencil Dwarf on Cloud for Democratizing Scientific Computing}}
\author{
{\rm Kun Li}\\
Microsoft Research
\and
{\rm Zhichun Li}\\
HPC Research Center, CAS
\and
{\rm Yuetao Chen}\\
Microsoft Research
\and
{\rm Zixuan Wang}\\
HPC Research Center, CAS
\and
{\rm Yiwei Zhang}\\
HPC Research Center, CAS
\and
{\rm Liang Yuan}\\
HPC Research Center, CAS
\and
{\rm Haipeng Jia}\\
HPC Research Center, CAS
\and
{\rm Yunquan Zhang}\\
HPC Research Center, CAS
\and
{\rm Ting Cao $^{\textrm{\Letter}}$ }\\
Microsoft Research
\and
{\rm Mao Yang}\\
Microsoft Research
}
 
\maketitle
\newcommand\blfootnote[1]{%
\begingroup
\renewcommand\thefootnote{}\footnote{#1}%
\addtocounter{footnote}{-1}%
\endgroup
}
\thispagestyle{empty}
\blfootnote{\hspace{-1em}$^{\textrm{\Letter}}$ Corresponding Author.} 
\blfootnote{\hspace{-1em} This work was completed in 2022.} 

\subsection*{Abstract}
{Stencil computation is one of the most important kernels in various scientific computing.}
{Nowadays, most Stencil-driven scientific computing} still relies heavily on supercomputers, suffering from expensive access, poor scalability, and duplicated optimizations.
{This paper proposes Tetris, the first system for high-performance Stencil on heterogeneous CPU+GPU, towards democratizing Stencil-driven scientific computing on Cloud. In Tetris, polymorphic tiling tetrominoes are first proposed to bridge different hardware architectures and various application contexts with a perfect spatial and temporal tessellation automatically. Tetris is contributed by three main components: (1) Underlying hardware characteristics are first captured to achieve a sophisticated Pattern Mapping by register-level tetrominoes; (2) An efficient Locality Enhancer is first presented for data reuse on spatial and temporal dimensions simultaneously by cache/SMEM-level tetrominoes; (3) A novel Concurrent Scheduler is first designed to exploit the full potential of on-cloud memory and computing power by memory-level tetrominoes.} Tetris is orthogonal to (and complements) the optimizations or deployments for a wide variety of emerging and legacy scientific computing applications. Results of thermal diffusion simulation demonstrate that the performance is improved by 29.6×, reducing time cost from day to hour, while preserving the original accuracy. 

\section{Introduction}

Over the past decades, scientific computing is prosperously developed to solve Grand Challenges~\cite{GrandChallenges}, such as solving genetic mysteries on Summit~\cite{grandchallenge}, exploring seismic simulation on Sunway Taihulight~\cite{fu20179}, and creating COVID-19 epidemic models on Fugaku~\cite{ando2021digital}, which utilize supercomputers to make significant progress.
{
}


\begin{figure*}
  \begin{center}
  \centering
  \includegraphics[width=0.97\textwidth]{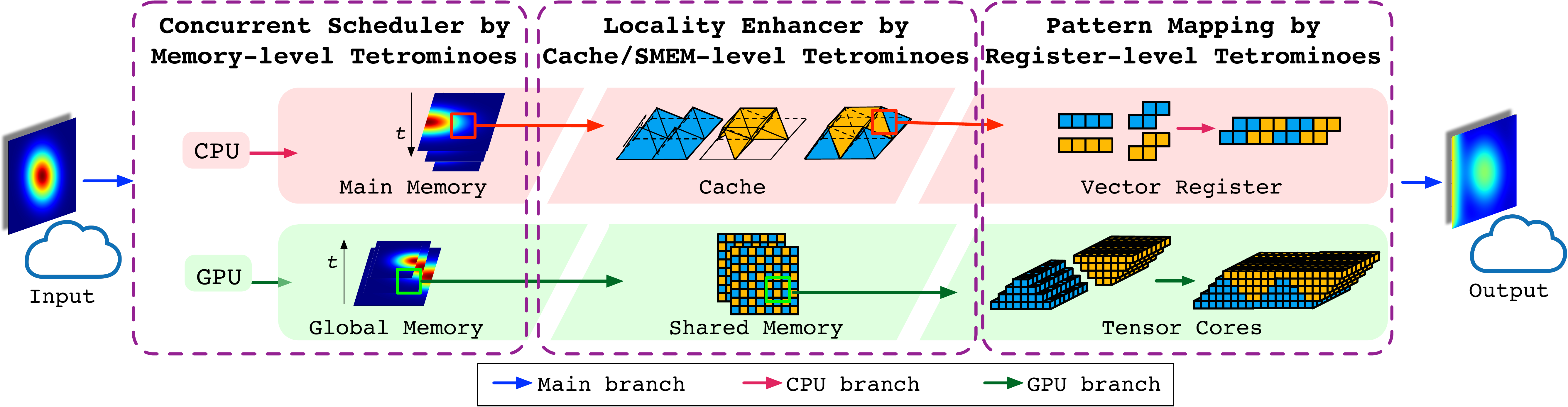}
  \caption{\label{fusion_overview}Tetris overview.}
  \end{center} 
\end{figure*} 

{Unfortunately, most scientific computing still relies heavily on traditional supercomputers~\cite{li2019openkmc,fu2017redesigning,fu20179}, and it hinders the democratization of science by three main obstacles. }
First, the access to a supercomputer is luxurious for users, and it is even prohibitively expensive to employ a full machine for {large-scale scientific computing}. 
Second, duplication of efforts is made on them by {engineers} case by case due to diverse supercomputer architectures.
Moreover, the scalability is poor as the assigned quota on a supercomputer is hard to adjust elastically.

Backed by advances in hardware and networking techniques, Cloud provides much higher performance scalability and cheaper access than traditional supercomputers. What is more, Cloud infrastructures, such as the ones from Microsoft Azure, Google Cloud, and Amazon Web Services~\cite{ogbole2021cloud,soh2020microsoft}, widely adopt unified CPU and GPU heterogeneous architecture, driving the prosperity of AI. Therefore, Cloud contributes a promising unified, high-performance, low-cost, and elastic infrastructure for {scientific computing}.

However, none of the work supports {scientific computing on Cloud for heterogeneous processors}. The traditional techniques on supercomputers cannot be easily applied to the Cloud either, because of the significant differences in architecture. {Although the AI surrogate models for scientific computing~\cite{jia2020pushing,cai2021physics,raissi2019physics} could potentially leverage the cloud resources, the introduced precision loss is not afforded by diverse applications with the demanding taste of accuracy} ~\cite{bailey2005high,bailey2012high,li2019openkmc}.

{Though the applications are diverse in scientific computing, analogy to AI, there are several common and performance-critical operations in scientific computing, named \textit{Dwarf}, defined by the Berkeley View~\cite{yuan2019tessellating}.} {As one of the seven computational Dwarfs, Stencil is ubiquitously involved in various scientific computing~\cite{datta2008stencil}, which lies at the heart of thermal diffusion ($\sim$100\%), earth system model ($>$90\%), and earthquake prediction model ($>$90\%), etc~\cite{li2021reducing,xiao2018communication,bluman1969general,heatformula,xushun2021,denzler2021casper,christen2012patus}.} 

To empower {scientific computing on Cloud, we take Stencil-driven scientific computing as a case study and} identify the following challenges.


\paragraph{Challenge \#1: Complicated Spatial and Temporal Data Dependencies.} 
{Scientific computing} is characterized by strong data dependencies in spatial and temporal dimensions since neighboring data are required to be collected to describe a phenomenon in a particular location and period of time, such as the computations of molecular motion, grid update, and interaction force among a variety of applications~\cite{li2019openkmc,yuan2019tessellating,fu20179,fu2017redesigning}. 
In {Stencil-driven scientific computing}, the exhaustively studied tiling is one of the most powerful techniques to improve data reuse and performance ~\cite{li2022efficient,yuan2019tessellating}. In general, they are designed for one type of memory on a specific hardware. However, there are different {on-Cloud memory types for CPU and GPU}, such as the registers and caches on the CPU and GPU, and the shared memory (SMEM) on GPU. One tiling shape cannot take full use of these resources, leading to severe performance degradation.

\paragraph{Challenge \#2: Unfriendly On-Cloud Hardware Architecture.}
High-performance facilities such as Vector Registers on CPUs and Tensor Cores on GPU suffer from over specialization, as only specific operations are supported on them. However, the operators in {scientific computing} are highly diverse, and they cannot reap the benefits of these facilities trivially.
On the one hand, though vectorization serves as an effective means of utilizing SIMD facilities on CPU, the data alignment conflicts are the main performance-limiting factors introduced by vectorization for {scientific computing}~\cite{henretty2011data,henretty2013compiler}, where the neighbors for a grid point appear in the same vector register but at different positions~\cite{li2022efficient}. On the other hand, Tensor Cores are specially designed for deep learning, and only simple matrix multiplication (MM) of specific size is supported~\cite{pisha2021accelerating,dakkak2019accelerating}. To the best of our knowledge, no previous work adopts Tensor Cores for {scientific computing} as it is prominently challenging to adapt abundant non-MM operations to the only one setting of FP64 MM operations on Tensor Cores~\cite{dakkak2019accelerating}.

\paragraph{Challenge \#3: Essential CPU and GPU Collaboration.}
Most applications on the Cloud use CPU or GPU only due to the large performance difference. However, {scientific computing} calls for the collaborated CPU and GPU execution. 
Firstly, {scientific computing} requires huge memory costs. For example, the memory needed per simulated atom is approximately 0.72KB and a large-scale kinetic Monte Carlo (KMC) simulation system is contributed by nearly hundred-billion atoms~\cite{li2019openkmc}! It is very costly to use GPU memory only, which calls for the use of CPU memory. Secondly, different from AI computations, {scientific computing} requires vast {FP64 operations}. GPU has a similar performance as the CPU for this kind of computation, which calls for the collaborated execution of CPU and GPU. Third, the collaboration of CPU and GPU introduces increased communication which complicates the computing process, leading to bubbles in the pipeline.

{\paragraph{\textit{Tetris.}} This paper proposes Tetris, the first system for high-performance Stencil on heterogeneous CPU+GPU, towards democratizing Stencil-driven scientific computing on Cloud.}


{The design of Tetris is based on two key observations.}

{First, changing data-update order would not break the application semantics. Specifically, Stencil aims to update data with restricted neighbors along the time dimension, while the updating order of data does not matter.}

{Second, creating a breakpoint would not destroy the computational logic. The dependencies of a point can be decomposed and computed partially first, and the remaining dependencies are collected at a postponed time.}



{Guided by these observations, the principal insight is: the data are not necessarily aligned spatially or temporally in a regular order; similarly, the calculation of the current point in a time step is also allowed with a brief pause for tessellating other calculations.}

  
{Based on this insight, the key design of Tetris is \textbf{\textit{tiling tetrominoes tessellation}}}. The high-level idea is inspired by a classic Tetris game of sliding blocks (named tetrominoes). In a Tetris game, different tetrominoes are just kept piling onto old ones first to fill the vacated spaces, {until they are aligned in a line. Analogously, with polymorphic tiling tetrominoes on different hardware hierarchies, Tetris carefully reorders the updating objects or marks a breakpoint on the partially-computed points, allowing a maximal updating is performed locally first while aligned spatially and temporally at a scheduled time point.}

{As a crucial thread in Tetris, tiling tetrominoes run through the whole memory hierarchy on heterogeneous processors of Cloud. Figure~\ref{fusion_overview} illustrates that Tetris consists of three main components: }

\paragraph{\textit{{Pattern Mapping by Register-level Tetrominoes}} }
Conventionally, only the workloads suitable for vectorization or MM operations are taken into account for offloading to specialized hardware units in existing work. 

Tetris captures the underlying hardware characteristics and {designs distinctive tetrominoes to perform a sophisticated pattern mapping from Stencil computation to vectorization and MM operations}.
Briefly, (1) Vector Skewed Swizzling is designed on Vector Registers, which 
utilizes skewed tetrominoes for building a conflict-free vectorized pipeline. It is the first design free of expensive cross-lane SIMD instructions (the CPU latency decreases from 3 to 1 per instruction~\cite{guide21url}). 
(2) Tensor Trapezoid Folding is proposed on Tensor Cores, and it employs stair tetrominoes to adapt non-MM operations in {Stencil} as a series of reduction and summation operations. It can leverage the computing power of Tensor Cores for MM meanwhile preserving high accuracy (FP64).

\paragraph{\textit{ {Locality Enhancer by Cache/SMEM-level Tetrominoes} } }
 


Instead of the conventional perspective that employs a single tiling shape to {improve data locality, Tetris achieves an efficient locality enhancer with polymorphic tiling tetrominoes for data reuse on spatial and temporal dimensions simultaneously.} 
{In the design of locality enhancer, the coarse-grained tetrominoes from memory could be tessellated as an aggregate unit of  tetrominoes on cache (or SMEM). Similarly, cache/SMEM-level tetrominoes can be decomposed as fine-grained tetrominoes to registers.
This flexible layout makes it possible for {scientific computing} adapted automatically to different architectures and application contexts simply by adjusting the number of data tetrominoes on cache (or SMEM).}

\paragraph{\textit{ {Concurrent Scheduler by Memory-level Tetrominoes} }}

Tetris proposes a novel concurrent scheduler designed specifically for {scientific computing} on the CPU and GPU. It identifies the implicit and explicit dependencies for {Stencil} workloads between CPU and GPU, {and a two-way partitioning is performed on input to obtain memory-level tetrominoes. } 
Concretely, (1) it explores a bidirectional design for squeezing memory consumption by maximizing memory savings on GPU while exploiting CPU memory for simulation.
(2) To preserve compute efficiency, a balanced {tetrominoes} partitioning strategy is designed to achieve orders-of-magnitude equal computation on CPU compared to GPU, preventing the CPU compute from becoming a performance bottleneck. (3) Moreover, the communication of {boundary tetrominoes} between CPU and GPU is significantly reduced to enable better scalability.

\paragraph{Results}
We have evaluated Tetris on varied classic benchmarks of Stencil Dwarf used for real-world applications. A thorough evaluation with Data Reorganization~\cite{10.1145/3126908.3126920}), Auto Vectorization~\cite{li2022efficient}, Pluto~\cite{bandishti2012tiling}, Folding~\cite{li2021reducing}, Brick~\cite{zhao2019exploiting} and AN5D~\cite{matsumura2020an5d} demonstrates that Tetris improves the performance by an overall of 4.4× and 2.8× on average compared to the state-of-the-arts (Data Reorganization and AN5D) and achieve a nearly linear scaling on Cloud. Furthermore, we employ Tetris for a simulation of thermal diffusion on a square copper plate where Stencil computations dominates overwhelmingly the whole simulation ($\sim$100\%) and it improves the performance of simulation by 29.6× \textbf{from day to  hour} while preserving the \textbf{original accuracy}.

\section{Background}


\subsection{{Stencil Dwarf}} 
\label{pde_derivation} 
{Dwarfs are defined as a series of classic algorithmic methods by the Berkeley View~\cite{asanovic2006landscape}, }
which earn the name by dragging down the performance in serving "Snow White" ({scientific computing}) due to {their performance-critical operations}. 

Listed as one of the most important Dwarfs, Stencil is extensively employed in various scientific and engineering applications, and arises as a principal class of floating-point kernels in science~\cite{asanovic2006landscape,asanovic2008parallel,10.1145/3126908.3126920}. 
Generally, Stencil contains a pre-defined pattern that updates each point in $d-$dimensional spatial grid iteratively along the time dimension. The value of one point at time $t$ is a weighted sum of itself and neighboring points at the previous time~\cite{10.1145/1989493.1989508}.

Listing~\ref{algrls} shows the Stencil of a regular Heat-2D (5 points) kernel, where c1 $\sim$ c5 are the accumulation coefficients. 
Here we use one of the heat equations as a case study to explore Heat-2D Stencil introduced in Listing~\ref{algrls}, and it models the physical transfer of heat in a region over time~\cite{heatformula}. Equation \ref{eq:heat1} is the standard form in three dimensions, where $U(t, x, y)$ represent the heat at a point $(x,y)$ at time $t$, and $\alpha$ is the thermal diffusivity.

\begin{algorithm}[b]   
    \caption{\label{algrls}Heat-2D Stencil.}  
    \begin{algorithmic}[1] 
    \small
    \Require mesh $A$, coefficient $c1 \sim c5$.
        \For{$t = 0 \to T$}
            \For{$i = 0 \to N_x$, $j = 0 \to N_y$} 
                \State  $A[(t + 1) \% 2][i][j]  = c1 \times A[t \% 2][i ][j] +$
                \Statex \qquad \qquad \quad  $ c2 \times A[t \% 2][i - 1][j] +  c3 \times A[t \% 2][i][j + 1] +$ 
                \Statex \qquad \qquad \quad   $   c4 \times A[t \% 2][i][j - 1] +  c5 \times A[t \% 2][i + 1][j]$
            \EndFor 
        \EndFor 
  \end{algorithmic}  
\end{algorithm}

\begin{equation}
\small
\label{eq:heat1}
\frac{\partial U(t, x, y)}{\partial t}=\alpha\left(\frac{\partial^2 U(t, x, y)}{\partial x^2}+\frac{\partial^2 U(t, x, y)}{\partial y^2}\right)
\end{equation}

To obtain a numerical algorithm performed on computers, we discretize the dimensions $x$, $y$, and $t$ into points spaced $\Delta x$, $\Delta y$, and $\Delta t$ apart, which transforms a point $(t, x, y)$ in continuous space as $(n \Delta t, i \Delta x, j \Delta y)$ in discretized space-time in Equation \ref{eq:heat2}:
\begin{equation}
\small
\label{eq:heat2}
    u^{n}_{i,j} \approx U(n \Delta t, i \Delta x, j \Delta y).
\end{equation}

With the discretization and simplification of Equation \ref{eq:heat1} by using CFL number $\mu$~\cite{heatformula}, 
the following update is yielded as:
\begin{equation}
\small
u_{i,j}^{n+1}=\left(1-4\mu\right)u_{i,j}^{n}+\mu\left(u_{i-1, j}^n+u_{i+1, j}^n+u_{i, j-1}^n+u_{i, j+1}^n\right),
\end{equation}
where Stencil is derived from heat equations in {scientific computing}.

\subsection{{Surrogate Models }}

\begin{figure}
  \begin{center}
  \centering
  \includegraphics[width=0.47\textwidth]{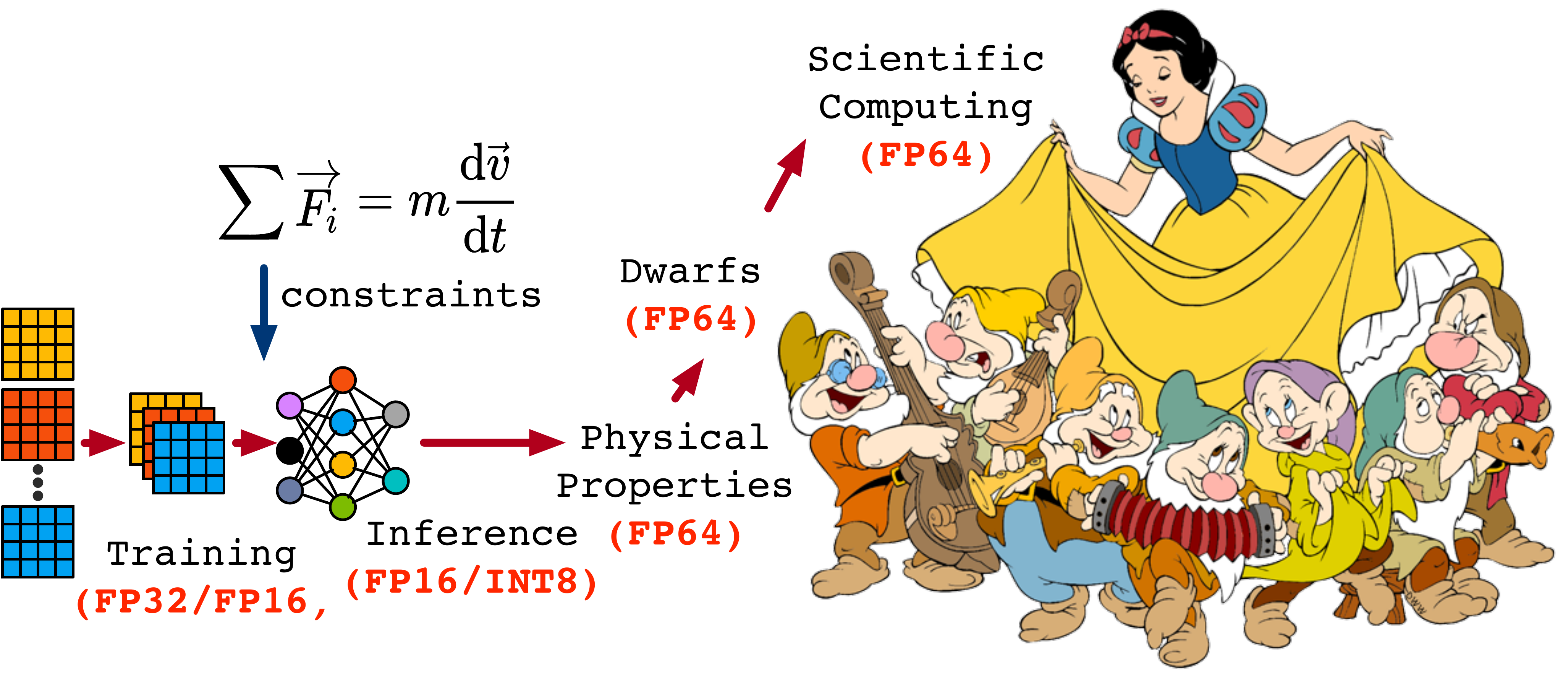}
  \caption{\label{old_fusion}AI surrogate models serving for {scientific computing}.}
  \end{center} 
\end{figure}

More recently, there has been considerable progress and promise in the use of AI surrogate models to complex calculations like first-principles functions~\cite{jia2020pushing}, which could utilize GPU to accelerate {scientific computing} and create potential possibilities for adapting to Cloud infrastructure. Albeit the use of AI surrogate models within {scientific computing} improves time-to-solution, most applications cannot utilize them due to the strict taste of accuracy and convergency, especially in the domains of weather forecast, galaxies collision, nuclear reactors, etc~\cite{bailey2005high,bailey2012high,li2021reducing}.  

Figure~\ref{old_fusion} provides a schematic representation of utilizing AI surrogate models for {scientific computing}~\cite{jia2020pushing}, which includes the three critical phases. 
First, AI surrogate models are designed and trained with simulation data. Once an appropriate trained model is built, it will replace partial calculations of physical properties. 
Second, the prepared physical properties are applied with Dwarfs for high-precision numerical simulation. 
At last, the numerical results produced by Dwarfs serve for "Snow White": the upper-level {scientific computing}, and a correction is promoted for potential biases or inaccuracies in trained models.

\section{{Pattern Mapping by Register-level Tetrominoes}}
With polymorphic register-level tetrominoes, Tetris designs a sophisticated {\textit{Pattern Mapping}} to make full use of specialized hardware units on heterogeneous architecture. 

\label{Sec:Template_Tetrominoes}

\subsection{Vector Skewed Swizzling} 
 
Existing vectorized optimizations for Stencil have mainly focused on either memory access or register permutation, aiming at decreasing transferred data volume and employing rich SIMD instructions respectively~\cite{li2021reducing}. To address the data alignment conflicts~\cite{li2022efficient} essentially, we perform a deep analysis of register architecture to find out the root cause of this challenge.

\paragraph{Vector Tetrominoes. }

\begin{figure}
  \begin{center}
  \centering
  \includegraphics[width=0.44\textwidth]{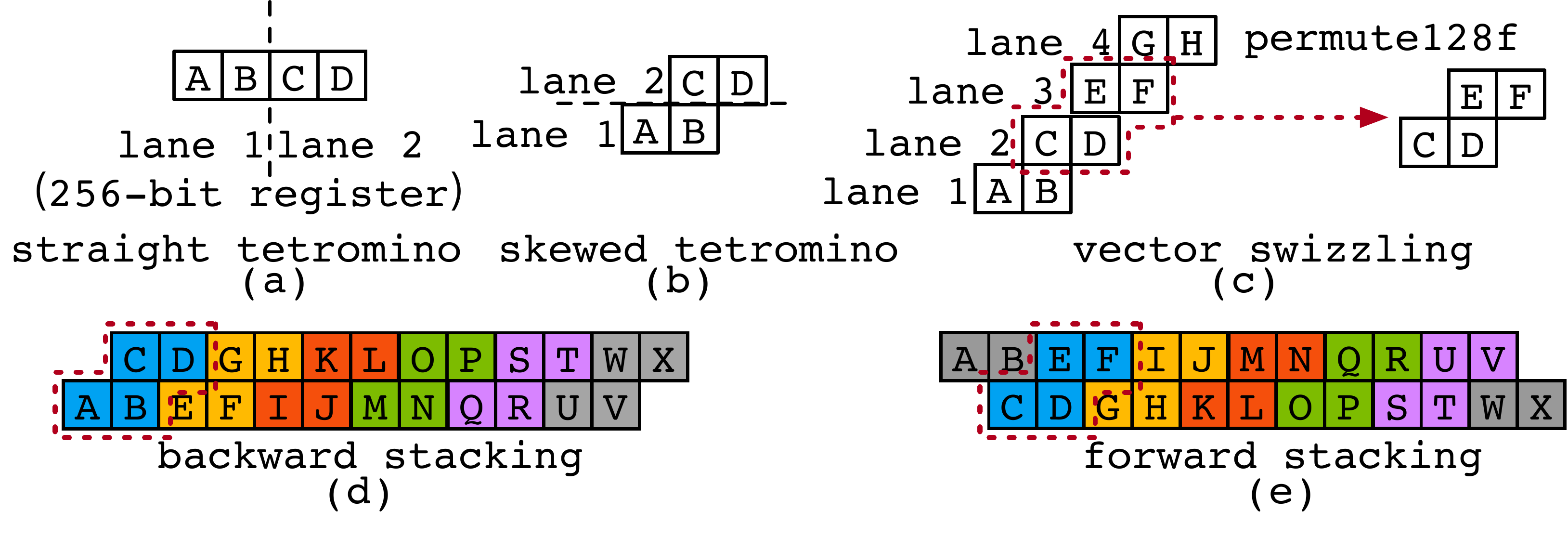}
  \caption{\label{cpu_layout}Skewed tetrominoes on CPU.}
  \end{center} 
\end{figure} 

Given a comprehensive view of registers in CPU, a 256-bit SIMD YMM register is actually promoted by two separate data lanes, where the lower 128-bits are aliased to a 128-bit XMM register~\cite{guide21url}. 
Taking FP64 computations into consideration, we fill up this 256-bit SIMD register with 4 elements initially as a straight tetromino.  
Since a 256-bit SIMD register is composed of 2 128-bit register lanes, a straight tetromino is also viewed as an assembly by 2 bars accordingly in Figure~\ref{cpu_layout}(a). 
To adapt this lane-based characteristic of registers, variants are derived from straight tetrominoes to skewed tetrominoes by employing the 128-bit lanes as basic processing units.

As shown in Figure~\ref{cpu_layout}(b), a skewed tetromino constituted by 4 squares could hold a complete vector, which also indicates a 256-bit SIMD register is configured with 4 $double$ type slots. Then we achieve \textit{forward stacking} in Figure~\ref{cpu_layout}(d) by manipulating a crowd of skewed tetrominoes to assemble pipeline forward, and the whole data space for a time step could be tessellated flawlessly in this way. 
In view of the lane-based characteristic of registers, an equivalent conversion principle from forward stacking to  \textit{backward stacking} in Figure~\ref{cpu_layout}(e) is also designed swimmingly by a cheap lane-based \verb=permute128f= instruction every two vectors illustrated in Figure~\ref{cpu_layout}(c). 

\paragraph{Skewed Swizzling.} 

\begin{figure}
  \begin{center}
  \centering
  \includegraphics[width=0.4\textwidth]{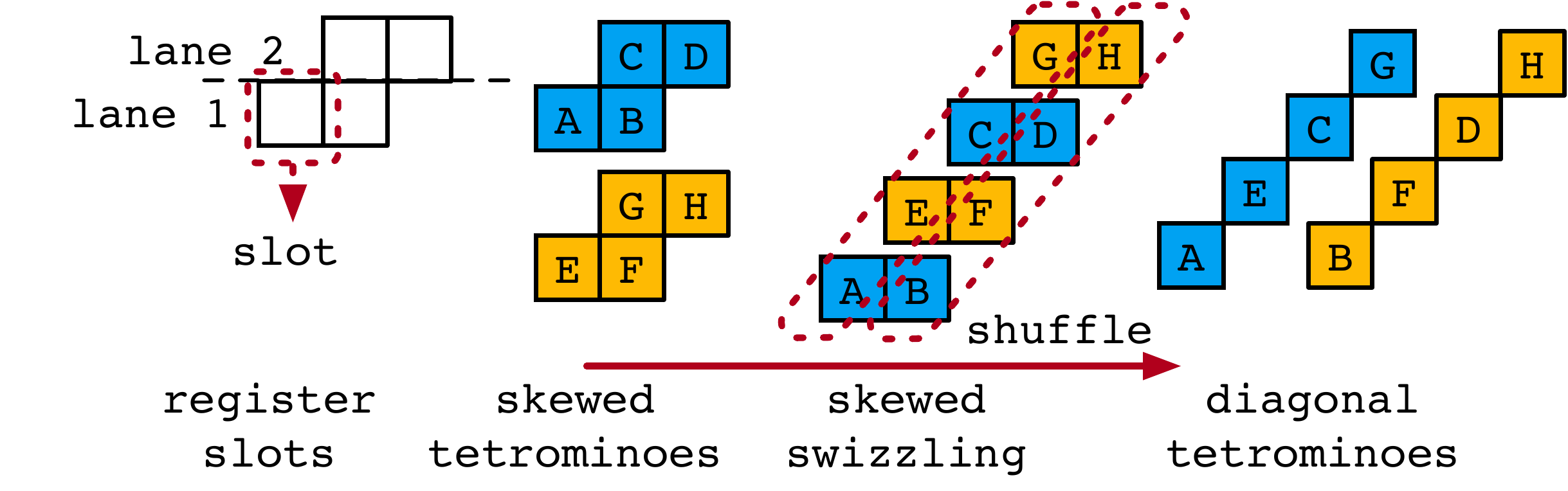}
  \caption{\label{cpu_swizzling}Skewed Swizzling in Vector Registers.}
  \end{center} 
\end{figure} 

The vectorization mechanism on CPU SIMD facilities is achieved by performing calculations on the slots of different registers vertically in one go, and each calculation flow on aligned slots is independent to each other. Thus, the key challenge to address the data alignment conflicts lies in an efficient adaption that remapping the adjacent elements to the same slot positions in different registers with minimal costs.

As illustrated preciously, lane-based operations are recommended for low-latency implementation when utilizing SIMD facilities on CPU. 
Figure~\ref{cpu_swizzling} shows the process of skewed swizzling for two tetromino registers, where elements at the same slot position of each 128-bit lane are swizzled with elements in the other register correspondingly. Specifically, the two skewed tetrominoes A, B, C, D and E, F, G, H are swizzled for A, E, C, G and B, F, D, H by unpacking and interleaving elements from the low half and high half of each 128-bit lane respectively, which only consumes 1 latency time by lane-based \verb=shuffle= instructions~\cite{granlund2012instruction}. Based on the efficient skewed swizzling, 
two skewed tetrominoes are swizzled as diagonal tetrominoes, where the data alignment conflicts are all well addressed with minimal cost.

\paragraph{Quadruple Pipelining.}

\begin{figure}
  \begin{center}
  \centering
  \includegraphics[width=0.49\textwidth]{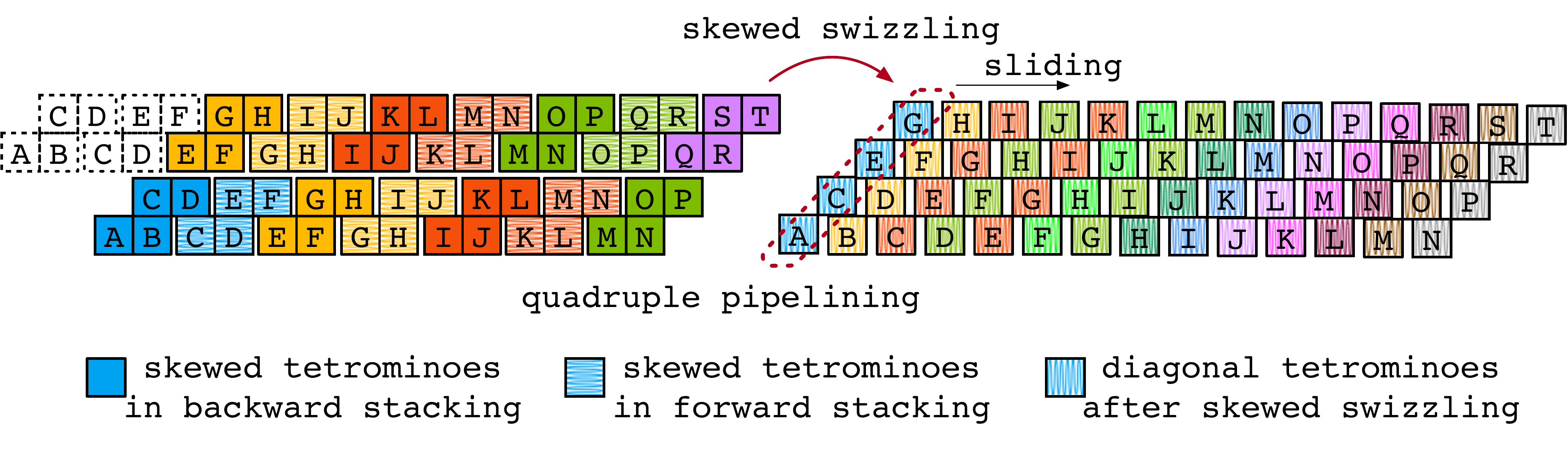}
  \caption{\label{cpu_pipeline}Quadruple Pipelining on Vector Registers.}
  \end{center} 
\end{figure}

Here we stack the forward and backward stacking in a well-designed principle to build a quadruple pipelining for a high-performance vectorized adaptation of Stencil Dwarf by Tetris.

Figure~\ref{cpu_pipeline} (a) exhibits the principle of quadruple pipelining, where the forward stacking is tessellated with backward stacking tightly. 
More specifically, each skewed tetromino in forward stacking is inserted at intervals in backward stacking orderly to build a dual pipelining first. Then the head tetromino in each stacking is removed, and similar operations are performed again to obtain a new staggered dual pipelining. At last, we stack two dual pipelinings vertically and still keep a skewed style, and the two skewed tetrominoes are further converted to diagonal tetrominoes with a simple skewed swizzling in Figure~\ref{cpu_pipeline} (b). With polymorphic tetrominoes, Tetris adapts Stencil Dwarf efficiently to CPU SIMD facilities by relieving data alignment conflicts in vectorization.  

\subsection{Tensor Trapezoid Folding}

\paragraph{Tensor Tetromino.} 

\begin{figure}
  \begin{center}
  \centering
  \includegraphics[width=0.48\textwidth]{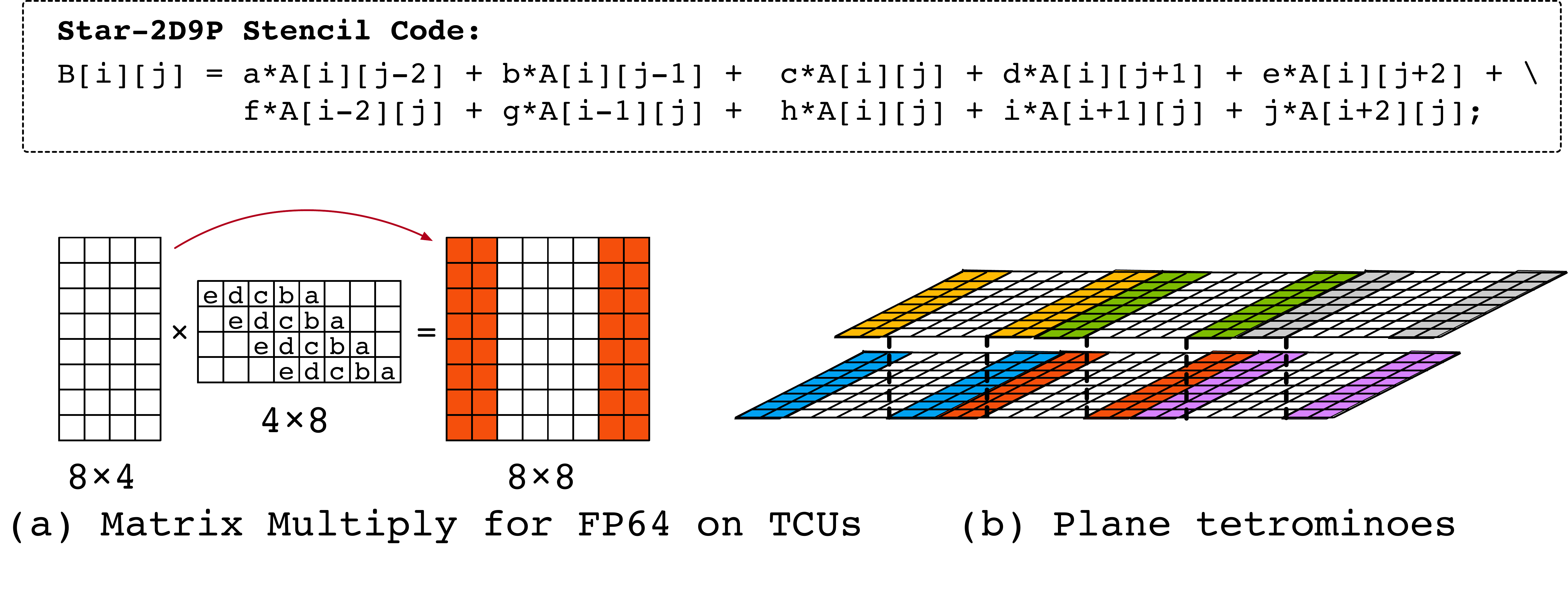}
  \caption{\label{gpu_tetromino}Plane tetrominoes on GPU.}
  \end{center} 
\end{figure} 

From NVIDIA Ampere GPU architecture like A100, the newly-introduced specialized hardware units known as NVIDIA’s Tensor Cores could offer a full range of precision to boost the performance of MM even with the highest accuracy needed (FP64).
Although Tensor Cores are prevalent and promising to provide an increase in performance, they suffer from over specialization as only MM on specific size of small matrices is supported. In particular, the size of supported matrix for FP64 precision is only constrained to one setting shown in Equation~\ref{eq:mma}:
\begin{equation}
\label{eq:mma}
   D_{m\times k} = A_{m\times n} \times B_{n\times k} + C_{m\times k},
\end{equation}
where m, n, k is 8, 4, 8 respectively.

Here a series of tensor tetrominoes are designed to achieve an efficient Stencil Dwarf adaptation. Concretely, at each step of Stencil, it applies the parameter kernels onto a portion of the elements in the specific shape, compute their element-wise products, and further sum the products together to update the value at this position. With this in mind, we could actually formulate the Stencil computation by using MM operations equivalently. 
Then Stencil is adapted as weighted reductions for a $8\times8$ plane tetromino.
At last, the adjacent plane tetrominoes are accumulated with overlaps for the final update at this position. 


\paragraph{Trapezoid Folding.} 

\begin{figure}
  \begin{center}
  \centering
  \includegraphics[width=0.48\textwidth]{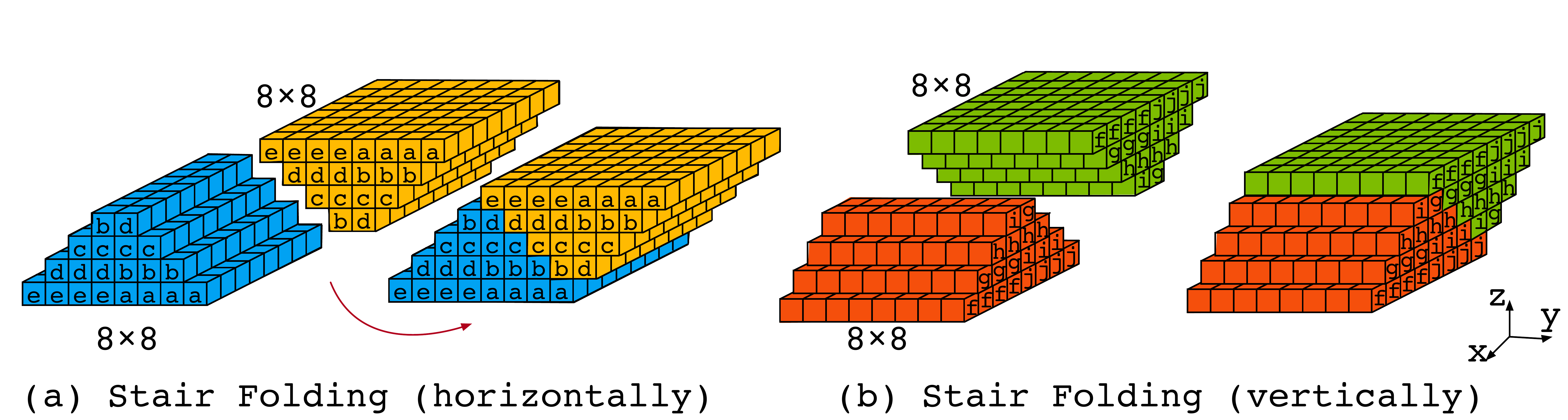}
  \caption{\label{gpu_folding}Tensor Trapezoid Folding on Tensor Cores.}
  \end{center} 
  \vspace{-0.4in}
\end{figure} 

A plane tetromino is an abstraction of the matrix $C$ described from a two-dimensional perspective. To make different weights counts, we upgrade the plane tetrominoes as stair tetrominoes from a three-dimensional perspective. Then a Trapezoid Folding strategy is presented to build a tessellation of data space by utilizing Stair tetrominoes as building blocks.

Figure~\ref{gpu_folding}(a) exhibits a sketch of Trapezoid Folding. 
Each position in a $xy-$plane tetromino rises prominently towards the $z$ dimension, building a symmetrical stair tetromino. 
The height is determined by the weights contributions from the parameter matrix $B$, and the positions in the same column are also of equal height due to the identical $B$ used in MM operations. 
To distinguish it more intuitively, the stair tetrominoes are inverted and colored alternatively, and then they are stacked to tessellate the whole data space. We call the
calculation along the third dimension with two stair tetrominoes a \textit{folding}. For example, a folding is performed first on the right part of stair tetrominoes A (blue) with the left part of adjacent stair tetrominoes B (yellow). Then the overlapped positions are all updated horizontally, where the heights are accumulated with 5 weights from all dependent neighbors. Similar operations are performed for obtaining a vertical folding in Figure~\ref{gpu_folding}(b).

\paragraph{Octuple Pipelining.} 

\begin{figure}
  \begin{center}
  \centering
  \includegraphics[width=0.48\textwidth]{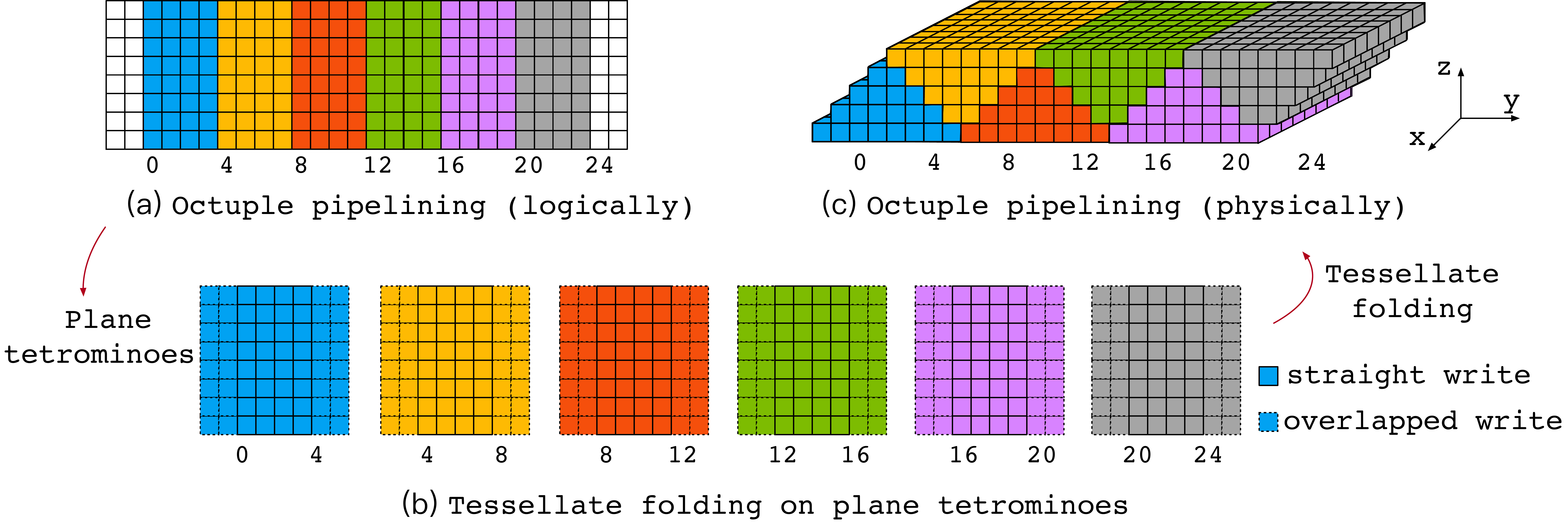}
  \caption{\label{gpu_pipeline}Octuple Pipelining on Tensor Cores (FP64).}
  \end{center} 
  \vspace{-0.2in}
\end{figure} 

Extended from vector to tensor, the tetrominoes defined on GPU are manipulated with a $8\times 8$ matrix in Tensor Core fragments. 
Since the elements of the same column in matrix $C$ are obtained by identical parameters in matrix $B$, the computations are performed along the $x$-direction horizontally, and each row represents a thread of calculation flow. 

Based on the efficient Trapezoid Folding and Checkerboard Blocking, we achieve Octuple Pipelining for Stencil in Figure~\ref{gpu_pipeline}.
First, the scientific data inputs on Checkerboard in shared memory are loaded into  fragments as a $8\times 4$ matrix  with different colors in Figure~\ref{gpu_pipeline} (a).
Then stair tetrominoes collecting dependencies of the adjacents are built and stacked by Trapezoid Folding to construct Octuple Pipelining. 
As shown in Figure~\ref{gpu_pipeline} (c), an efficient Octuple Pipelining is built to tessellate the whole data space, which achieves a high-precision adaptation from Stencil Dwarf to MM operations on Tensor Cores.

\section{{Locality Enhancer by Cache/SMEM-level Tetrominoes}}
\label{sec:Tiling_Generator}
{
In this subsection, we will introduce the \textit{Locality Enhancer} by tetrominoes of the middle layer (cache on CPU and shared memory on GPU) in memory hierarchy, which enhances the data reuse spatially and temporally meanwhile matching other tetrominoes in top and bottom memory hierarchies. }

\subsection{Tessellate Tiling} 
\begin{figure}
  \begin{center}
  \centering
  \includegraphics[width=0.48\textwidth]{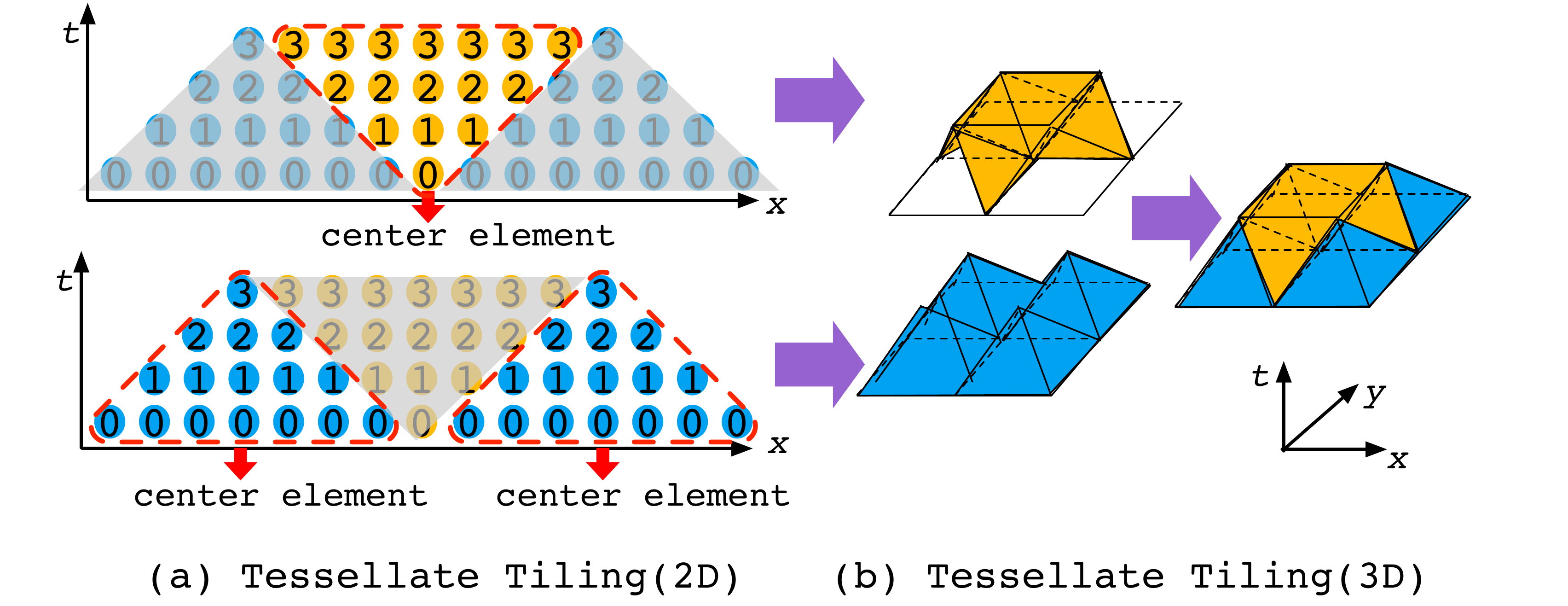}
  \caption{\label{cpu_tiling} {Tessellate Tiling ($T_b=3$). Triangle tetrominoes are tessellated in 2D and upgrade as tetrahedron tetrominoes in 3D.}}
  \end{center}  
\end{figure} 
{
Since modern high-performance CPU is generally configured with multiple cores, the data space can be tessellated spatially to enable fully parallelism such that all tetrominoes between synchronizations can execute concurrently without redundant computation. To develop more data reuse with fine optimizations, temporal tiling is then also presented by stretching the tetrominoes along the time dimension additionally.}

Typically, the working sets are many orders of magnitude larger than a processor’s cache memory. 
Let $T_d\{m,n\}$ be the data transfer time through the memory hierarchy $m$ to $n$, and we use this shorthand notation for expressing data transfer time in each iteration (e.g., $T_d\{L2,Reg\}=T_{L2L1}+T_{L1Reg}$ when the data is in the L2 cache). 
Since all grid cells are updated once before the next update sweep starts, a bad data locality is common to Stencil Dwarf with $T_d\{Mem,Reg\}=T_{MemL3}+T_{L3L2}+T_{L2L1}+T_{L1Reg}$, where $T_{MemL3}\gg T_{L3L2}>T_{L2L1}>T_{L1Reg}$.

\paragraph{Spatial Tiling.} 
Tessellate Tiling is used first by employing shaped tetrominoes to reduce $T_d\{Mem,Cache\}$, which dominates a major cost of data transfer time~\cite{yuan2019tessellating,li2021reducing}.
It can be viewed as a tessellation in iteration space by utilizing shaped tetrominoes. Figure~\ref{cpu_tiling} illustrates the tiling strategy for a two-dimensional Stencil computation with $T_b=3$ (temporal tiling size), where the timestamp of each element along the time dimension is annotated on it. 
Triangle tetrominoes are updated first, where the center element is updated three steps and its neighbors are updated fewer steps proportional to the distance with the center element.
Then two half parts from adjacent triangle tetrominoes constitute new inverted triangle tetrominoes, and identical operations are performed to make all elements updated with the same steps.
From the perspective of front view, The iteration space is tessellated by triangle tetrominoes and inverted triangle tetrominoes in alternative stages. 

\paragraph{Temporal Tiling.} 
As shown in Figure~\ref{cpu_tiling}, concurrent execution is processed by two stages over the same steps temporally with spatial tessellation completed, and the updated step of a specific element in a tetrominoes is proportional to its distance with the center element. Taking updated steps into account, the 2D iteration space in Figure~\ref{cpu_tiling} rises straight from the ground. At this point, 
the 2D triangle tetrominoes are also expanded as 3D tetrahedron tetrominoes, and two types of them tessellate the whole data space along the spatial and temporal dimensions.

With the tessellate tiling strategy, concurrent execution for different tiles is enabled over a given time range without redundant computation~\cite{yuan2019tessellating}, and a high in-memory flops/byte ratio is achieved with more data reused.

\subsection{Checkerboard Blocking}  
{An efficiently optimized GPU code must allocate appropriate GPU memory resources, especially on shared memory, to improve data locality. Most code generators automatically determine this resource assignment (or mapping), but without considering the balance on resource limits of the underlying GPU device and the requirements of applications. Taking the warp size (32 threads) and SMEM capacity into account, Checkerboard Blocking is carefully designed to provide an enhanced locality manually.  }
\begin{figure}
  \begin{center}
  \centering
  \includegraphics[width=0.45\textwidth]{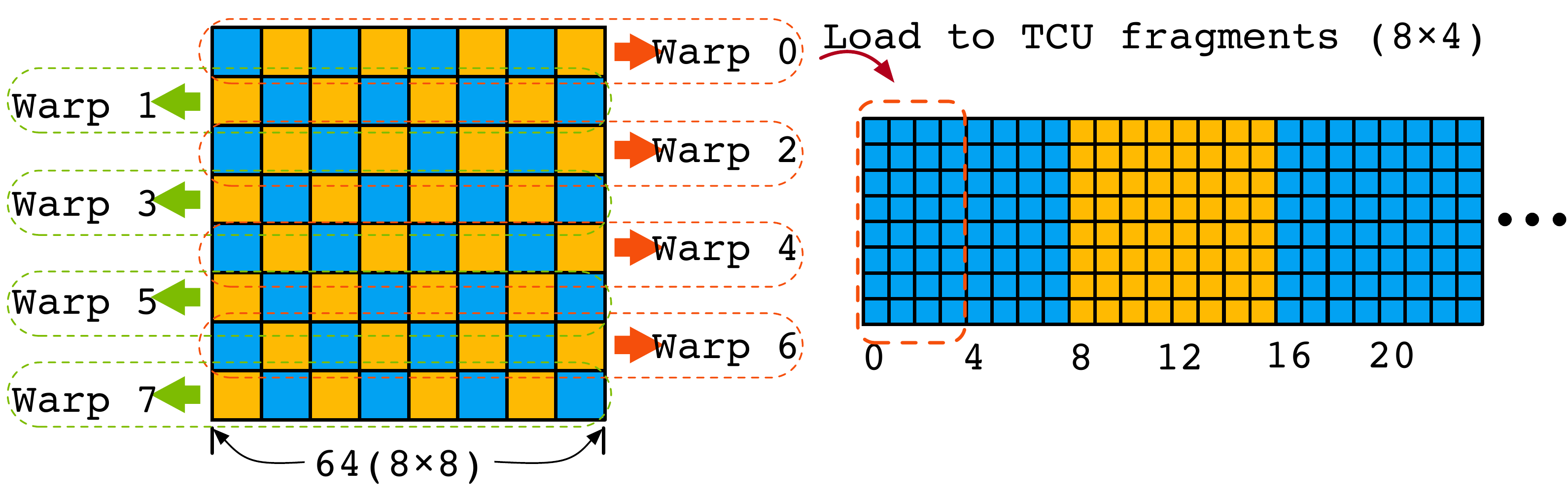}
  \caption{\label{gpu_blocking}Checkerboard Blocking on shared memory. }
  \end{center} 
  \vspace{-0.2in}
\end{figure} 
 
\paragraph{Spatial Blocking.} 
As a bridge for connecting global memory and registers, the shared memory is utilized as the stored buffer to guarantee the coalesced global memory access, where the computed results from registers will also be stored back to it first.
As shown in Figure~\ref{gpu_blocking}, Checkerboard Blocking strategy is proposed for utilizing the shared memory efficiently. To achieve a high parallelism, the shared memory in a GPU block is abstracted as an $8\times 8$ checkerboard that is criss-crossed by two types of square tetrominoes. Each row on the checkerboard is computed by a warp, and the threads in a warp are responsible for each concrete square tetromino. With Checkerboard Blocking, the data space on shared memory is occupied spatially without leaving any memory space unexploited or computing threads idle, and the coordinated
data exchange could be performed with coalesced writes to global memory.

\paragraph{Conflict-free Blocking.}
It is required to address two major different conflicts on shared memory for unlocking a higher performance. One is the bank conflict triggered by hardware mechanism, and the other is block conflict introduced by inter-block compute dependencies. 
Instead of padding the shared memory buffer with additional columns, Checkerboard Blocking reduces the read and write bank conflicts entirely by a precise control of thread access. Specifically, each square tetromino on checkerboard is further decomposed into an $8\times 8$ grid, and a grid point represents a real data. Then the half part of tetromino with $8\times 4$ grid points is loaded into a fragment on Tensor Cores, where the data and threads are mapped with a biunique correspondence without any conflict. As for block conflicts, Checkerboard Blocking stagger the square tetrominoes with two colors in Figure~\ref{gpu_blocking}, and the updates are performed alternatively by warp scheduling and thread controlling precisely  to avoid inter-block compute dependencies.

\section{{Concurrent Scheduler by Memory-level Tetrominoes}}

\begin{figure}
  \begin{center}
  \centering
  \includegraphics[width=0.48\textwidth]{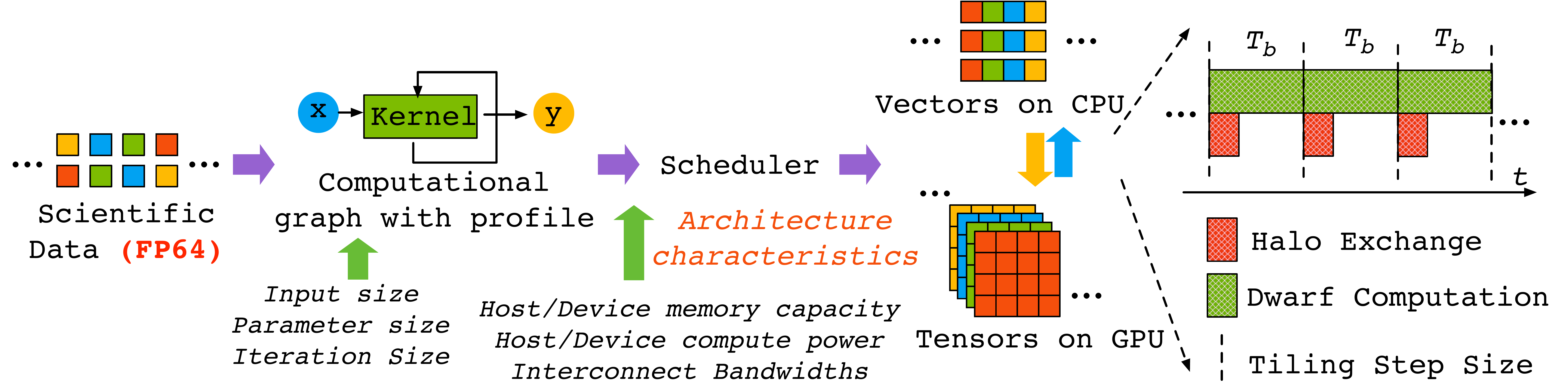}
  \caption{\label{tetris_scheduler}Concurrent heterogeneous scheduler.}
  \end{center} 
  \vspace{-0.2in}
\end{figure} 

{Tetris identifies the implicit and explicit dependencies between CPU and GPU, and a two-way partitioning is then performed on input as memory-level tetrominoes. Towards an efficient \textit{Concurrent Scheduler} on heterogeneous processors, the following techniques are proposed on memory-level tetrominoes.}

\subsection{Bidirectional Memory Squeezing} 

Memory consumption is still a tough challenge for {scientific computing}, especially in large-scale high-precision simulation~\cite{ren2021zero,li2019openkmc}.
Unlike previous work only utilizing CPU memory for high-precision simulation, Bidirectional Memory Squeezing in Tetris is designed to enable more efficient simulation by sharing memory resources on CPU and GPU. 

The {scientific computing} workloads can be represented as a directed graph of data and computation. Figure~\ref{tetris_scheduler} shows a Stencil Dwarf graph, where the output in the current step will be used as input for the next step. Since the spatial dependencies are much weaker than temporal dependencies, the offload strategy between GPU and CPU can then be abstracted by employing a two-way partitioning of input, such that
tetrominoes in a partition would be executed and stored on that compute worker to explore more memory quota. It is also worth noting that the strategy is achieved with little offload cost for more computing and storage benefits.

Bidirectional Memory Squeezing is also a bidirectional design for squeezing memory consumption. Once the GPU memory is fully occupied, the remaining part left on CPU is still well-addressed since both CPU and GPU provide memory and compute for better efficiency. When multiple GPUs are available, Bidirectional Memory Squeezing could be extended similarly to offer excellent scalability in even larger {scientific computing} applications.

\subsection{Auto-tuning Computation Scheduling} 
Tetris treats {scientific computing} applications as decomposable workloads since tetrominoes are coordinated explicitly by it, with each type of worker (CPU or GPU) executing a subset of the tetrominoes. Like with any distributed system, the steady state throughput of the whole heterogeneous architectures is determined crucially by the slowest worker. Having CPU and GPU process subtask at vastly different throughput can lead to bubbles in
the computing pipeline, starving the faster worker of subtasks to work on and
resulting in resource under-utilization.

Tetris sticks to the fact that Stencil Dwarf exhibits little variance in computation time across the same size of inputs. As shown in Figure~\ref{tetris_scheduler}, in the startup phase, the input stage admits enough tetrominoes to keep each worker full in steady state.
Then the computation time taken by the first iteration, the input size
of scientific data, the size of Dwarf parameters and the number of iterations are recorded as part of a profile initialization. This profile is employed as the input to the  scheduler for performing a balanced partition on a heterogeneous platform.
Apart from the constraints of software, the scheduler is also architecture-aware in a design of taking into account other hardware characteristics, such as host/device memory capacity, host/device compute power, and the interconnect bandwidths, and it computes: (1) a partitioning on the input into workers, (2) the estimated communication or replication volume between workers, and (3) optimal number of in-flight tetrominoes to keep the computing pipeline busy.

Intuitively, this process finds the optimal partitioning on CPU and GPU and can also be extended easily across clusters. Based on the partitioning strategy, Tetris’s scheduler dispatches a balanced workload to each worker on the main memory/global memory level.

\subsection{Minimized Communication Cost} 

\paragraph{Less Communication Volume.} Most of the existing work employing GPU accelerators requires a busy data transfers between CPU and GPU, as the CPU only plays a role of stream controlling and data is updated between CPU and GPU round by round. 
Tetris can communicate far less than this. Instead of having to perform a memory turnover periodically, each worker first deals with the halo region introduced by balanced computation scheduling, and has to communicate only this small subset of the working set. 
\paragraph{More Communication Overlap.} Despite using a highly reduced halo exchange, the communication interlude can still become a bottleneck that disturbs the regular computing pipeline. For such limited cases, we separate the streams of computation and communication carefully for a parallel execution, and address halo updates first in each working set. As shown in Figure~\ref{tetris_scheduler}, with the halo region updated, the exchange is performed immediately, which overlaps computation and communication to hide the halo exchange overhead. 

\paragraph{Centralized Communication Launch.} For the communication part, we collect the halo exchanges for $T_b$ steps and then only conduct one centralized communication instead of conducting $T_b$ small communications.
The reason is analyzed below. The time for communication could be formulated by $k\times (\alpha + n_b \times \beta)$, where $k$ is the number of messages; $\alpha$ is launch latency; $1/\beta$ is the bandwidth; $n_b$ is the number of bytes per message. In practice, since we have $\alpha >> \beta$, an intact big message is much cheaper than $k$ split small messages due to $k\times (\alpha+n_b\beta)>>(\alpha+k\times n_b\times \beta)$. Thus, centralized communication launch are proposed to further decrease communication cost for $T_b$ steps. 

\section{Evaluation}

\subsection{Experimental Setup}

\begin{table}[b] \small 
  \caption{Configuration for Stencil benchmarks}\label{parameters}
  \renewcommand\tabcolsep{1.5pt} 
  \centering\begin{tabular}{lccr}
  \toprule
  Type & Pts & Problem Size & Blocking Size\\ \hline
  Heat-1D&3    & 10,000,000$\times$100,000 & 2,000$\times$1,000  \\
  Star-1D5P&5    & 10,000,000$\times$100,000 & 2,000$\times$500   \\
  Heat-2D&5    & 10,000$\times$10,000$\times$10,000 &  200$\times$200$\times$50                \\
  Star-2D9P&9    & 10,000$\times$10,000$\times$10,000 &  200$\times$200$\times$50                \\
  Box-2D9P&9    & 10,000$\times$10,000$\times$10,000 & 2,000$\times$500   \\
  Box-2D25P&25    & 10,000 $\times$10,000$\times$10,000 & 120$\times$128$\times$60                \\
  Heat-3D&7    &  1,024$\times$1,024$\times$1,024$\times$1,000   &20$\times$20$\times$10                     \\
  Box-3D27P&27   &   1,024$\times$1,024$\times$1,024$\times$1,000           &  20$\times$20$\times$10                    \\ \bottomrule
  \end{tabular}
\end{table}  
\paragraph{Machine} Experimental results presented in this paper are obtained by using a high-performance node on Microsoft Azure, which is composed of an AMD EPYC 7V13 processor with 24 physical cores of 2.45 GHz clock speed. It features the AVX2 SIMD instruction set, and each core contains a 768 KB private L1 cache, a 12 MB private L2 cache, and a unified 96 MB L3 cache. DDR4 DRAM and 8 memory channels are supported for a total 216 GB memory, and it yields a peak memory bandwidth of 409.6 GBps. An Nvidia A100 GPU is also configured with 80 GB memory, and the memory bandwidth could reach 1,935 GB/s. The most characteristic Tensor Cores with 108 SMs deliver a peak FP64 throughput of 19.5 TFLOPS, which is 2.5x that of Nvidia V100. 

\paragraph{State-of-the-arts} 
Although recent studies on Stencil Dwarf only exhibit their own absolute performance without comparison in experiments~\cite{matsumura2020an5d,zhao2019exploiting,ahmad2021fast}, we reproduce artifacts exhaustively and compare the performance of different state-of-the-arts for a comprehensive analysis.
Two classic vectorization methods (Auto Vectorization~\cite{li2022efficient} and Data Reorganization~\cite{10.1145/3126908.3126920}) are employed first as a standard baseline on CPUs. Then the newly-related state-of-the-arts (Tessellation~\cite{yuan2019tessellating} and Folding~\cite{li2021reducing}) are adopted for further comparison. At last, the highly-optimized work with GPU support, Brick~\cite{zhao2019exploiting} and AN5D~\cite{matsumura2020an5d}, are also evaluated thoroughly.
We utilize the OpenMP pragma \textit{parallel for} on all methods for scalability experiments.

\paragraph{Benchmarks} We use a variety of Stencil kernels with different shapes as benchmarks. Details are listed in Table \ref{parameters}, which consists of five star kernels (Heat-1D, 1D5P, Heat-2D, Star-2D9P, and Heat-3D) and three box kernels (Box-2D9P, Box-2D25P, and Box-3D27P) drawn from~\cite{10.1145/3126908.3126920,li2021reducing,bandishti2012tiling}.  
\paragraph{Programming}
Provided with a mathematical look into various benchmarks, we abstract the application skeletons as a semi-automatic code generator with Python wrapper. For programming on CPU, we use C/C++ for Tessellate Tiling and AVX2 SIMD instruction set for Vector Skewed Swizzling~\cite{guide21url}. Then Checkerboard Blocking is programmed with CUDA C/C++ and Tensor Trapezoid Folding is coded in a manner of MM by Warp Matrix Multiply and Accumulate (WMMA) API~\cite{CUDAToolkit}.

\paragraph{Metrics}
Most work on Stencil Dwarf (e.g., Tessellation~\cite{yuan2019tessellating,10.1145/3126908.3126920}, Pluto~\cite{bandishti2012tiling}, Folding~\cite{li2022efficient,li2021reducing}, etc.) exhibit results in terms of arithmetic performance (Stencils/s). In this work, we also adopt the metric of stencils per second (Stencils/s) defined in Equation~\ref{eq:stencilss} for measuring the performance. Here, $N_x$, $N_y$, $N_z$ are the problem size for each spatial dimension; $T$ is the iteration item in temporal dimension; $time$ is the total execution time. 
\begin{equation}
\label{eq:stencilss}
   \textbf{ stencils per second}=\frac{N_x\cdot N_y\cdot N_z}{time} \times T
\end{equation}

\subsection{Performance Breakdown}
\begin{figure}
  \begin{center}
  \centering
  \includegraphics[width=0.48\textwidth]{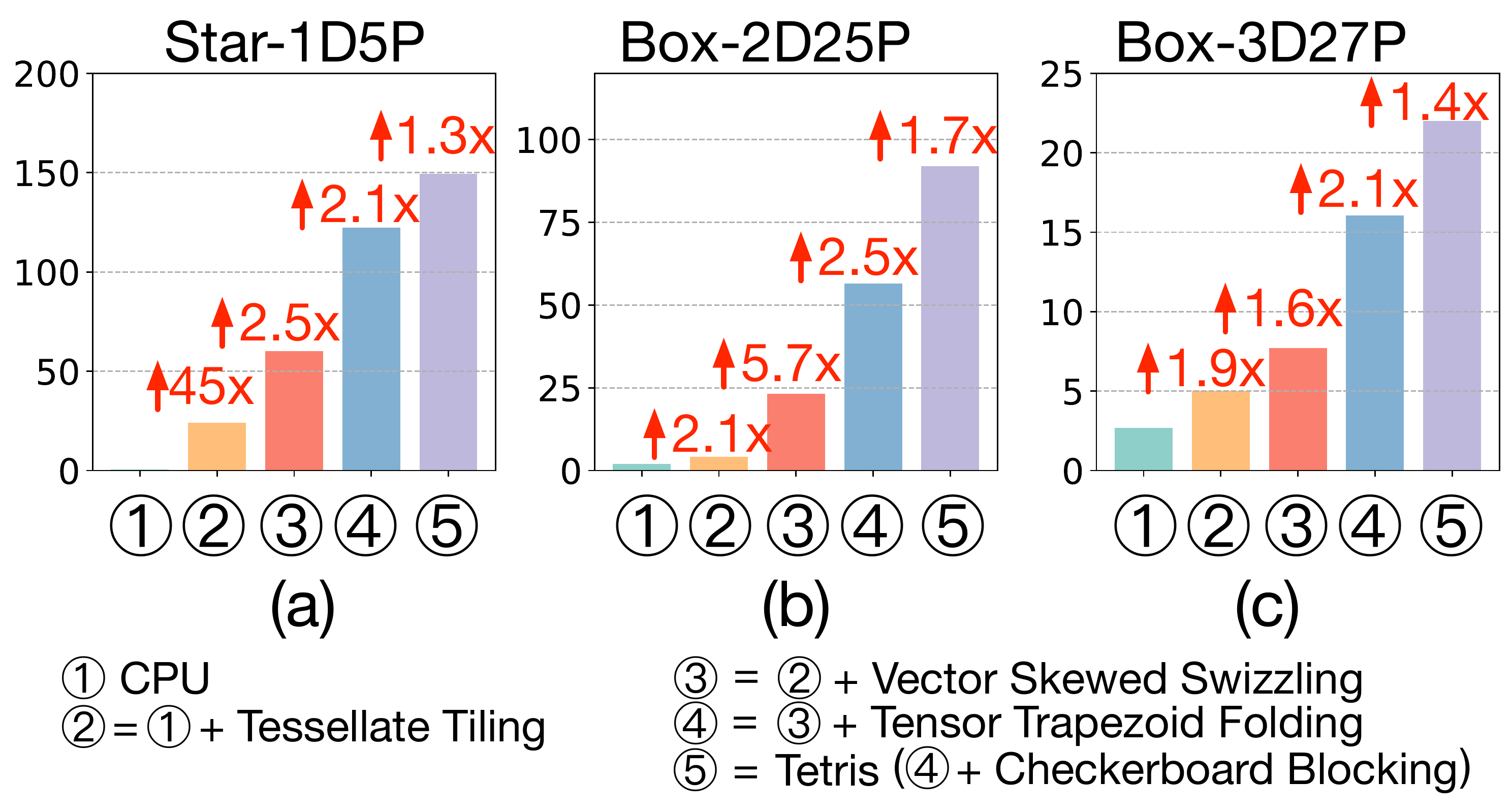}
  \caption{\label{breakdown}{Performance breakdown of Tetris.}}
  \end{center} 
  \vspace{-0.2in}
\end{figure}  
In this subsection, we investigate how the performance is improved progressively with different optimizations by Tetris.

Figure~\ref{breakdown} shows the performance breakdown of Tetris on three representative benchmarks (Star-1D5P, Box-2D25P, and Box-3D27P) as they contain more complex data access patterns and pose greater performance demands of Tetris. 


As can be seen from Figure~\ref{breakdown}, optimizations on CPU in Tetris are first investigated on three benchmarks. Tessellate Tiling enhances the data reuse in CPU cache effectively and even provide a 45x speedup considerably in Star-1D5P benchmark. Next, Vector Skewed Swizzling utilizing the high-performance SIMD facilities in CPU is performed and apparently outperforms the previous version in all experiments. Since Vector Skewed Swizzling reduces redundant loading costs and minimizes the intra- \& inter- register operations, it could relieve data alignment conflicts in vectorization efficiently and provide significant speedups from 1.6x to 5.7x. 
Here we have shown that a nerfed Tetris with multicore CPU could reach 112.5x, 12.0x, 3.1x speedups respectively on these representative benchmarks.

Then GPU is introduced in Tetris for further performance improvements. With the deep exploitation of Tensor Cores, all benchmarks achieve a minimum of doubled performance compared to the nerfed Tetris on CPU. 
Like CPU cache, shared memory provides more efficient data retrieves but limited capacity between local register and global memory. After utilizing shared memory by Checkerboard Blocking, 1.3x-1.7x speedups are obtained compared to the previous version. Up to this point, Tetris has accumulatively achieved speedups of 307.1x, 50.9x, 9.0x respectively, showing significant performance benefits brought by Tetris on multidimensional benchmarks.



\subsection{State-of-the-art Comparison \label{eva:comp}}

\begin{figure*}
  \begin{center}
  \centering
  \includegraphics[width=0.98\textwidth]{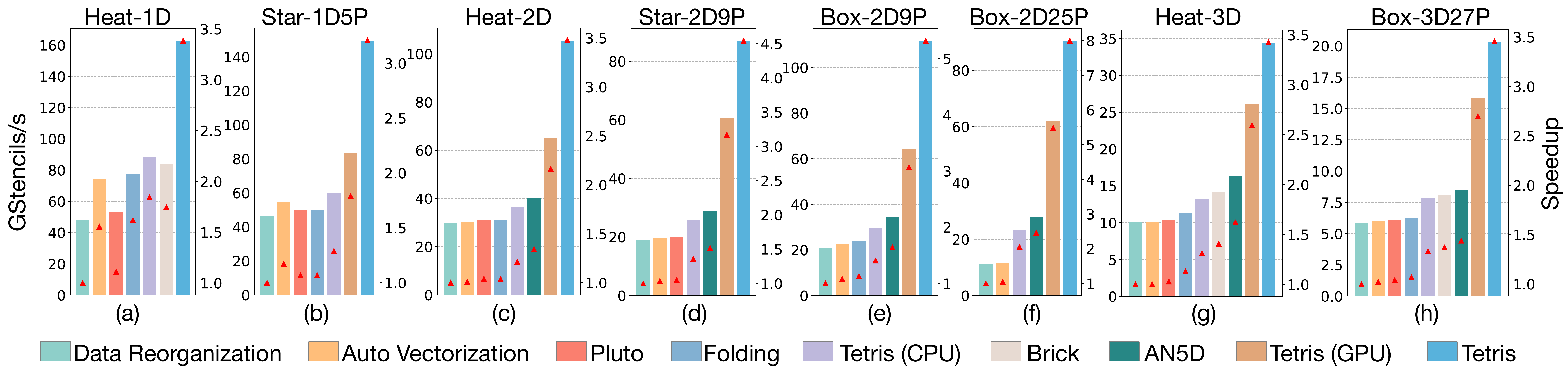}
  \caption{\label{state-of-the-art}{Performance and speedup comparison of state-of-the-arts and Tetris. The speedups of each group are compared to the lowest base which is annotated with the triangles by default value of 1.0.} }
  \end{center}
  \vspace{-0.2in}
\end{figure*}

\begin{table}[hbtp]
  \centering \small  
  \begin{threeparttable}
  \renewcommand\tabcolsep{0.1pt} 
  \caption{\label{tab:benchcomp}{Technical overview of state-of-the-arts and Tetris}}
  \begin{tabular}{lccr}
  \toprule {Methods}&Tiling&  Pipeling  &{Architectures} \\  \midrule 
    Data Reorg.~\cite{10.1145/3126908.3126920} & Split  & Data Reorg. &  C(VR)$^1$ \\
    Auto Vec.~\cite{li2022efficient}  & Split  &Auto Vec.&  C(VR) \\
   Pluto~\cite{bandishti2012tiling}& Diamond~\cite{bondhugula2008practical}&AutoVec.   &  C(VR) \\ 
   Folding~\cite{li2021reducing}&Polyhedral&  Folding  & C(VR)  \\
   Tetris (CPU) &Tessellate &  Skewed$^2$ &C(VR)  \\
   Brick~\cite{zhao2019exploiting}&  Brick & Scatter  & G(CCU)  \\
   AN5D~\cite{matsumura2020an5d} & Temporal & - &G(CCU) \\
   Tetris (GPU)& Checkerboard & Trapezoid &G(TCU)  \\
   Tetris &Polymorphic & Template &C(VR) + G(TCU)  \\ 
  \bottomrule \end{tabular} \footnotesize
  \begin{tablenotes}
  \item[1] For better clarity, CPU, GPU, Vector Register, CUDA Core Units and Tensor Core Units are abbreviated with C, G, VR, CCU and TCU respectively. 
  \item[2] Notice we use a keyword as an abbreviation for some algorithms. 
  \end{tablenotes}
  \end{threeparttable}
  \vspace{-0.2in}
\end{table} 

In this subsection, we present the experiments that exhibit the benefits of Tetris with various state-of-the-arts on different benchmarks, and the technical overview of them are listed in Table~\ref{tab:benchcomp}.

As shown in Figure~\ref{state-of-the-art}, taking all benchmarks with AVX2 instructions on CPU into account, remarkable performance improvements are observed from Tetris (CPU). Compared with the newly-released state-of-the-art Folding, Tetris (CPU) improves the overall performance sustainably by an average of 21\% and a maximum of 25\% in Box-2D9P and Box-3D27P benchmarks, which demonstrates that our optimization on CPU architecture could provide a significant benefit especially in complex pattern compared to the referenced work.

Here, to evaluate whether a nerfed Tetris only run on GPU can achieve a competitive performance, we additionally employ another two state-of-the-arts for a further comparison, where AN5D is a highly-optimized work using CUDA Cores. Apparently, Figure~\ref{state-of-the-art} shows Tetris (GPU) achieves the best for all benchmarks  even compared to the previous CPU baselines. Compared to AN5D, Tetris (GPU) achieves an overall 1.9x speedup on average and even reach a maximum 2.2x speedup in Figure~\ref{state-of-the-art}(f)! The reason is that, the Dwarf computation on Tensor Core is adapted efficiently by Tensor Trapezoid Folding, and Checkerboard Blocking make full use of shared memory for data reuse especially more friendly to complex benchmarks.

When Tetris is performed on CPU and GPU, the performance potential is fully expressed and it runs rings around all state-of-the-arts. We still observe that the Tetris performance is arithmetically close to the sum of two nerfed versions (Tetris (CPU) and Tetris (GPU)) for all benchmarks. This is because the cost of communication between CPU and GPU is greatly reduced and overlapped with computation, and the auto-tuning scheduling mechanism provides a balanced task partition, which make the utmost of computing power on Cloud. Compared to Data Reorganization, Tetris improves the overall performance by an average of 4.4x and a maximum of 8.1x in Box-2D25P.


\subsection{Scalability Evaluation}
\begin{figure}
  \begin{center}
  \centering
  \includegraphics[width=0.48\textwidth]{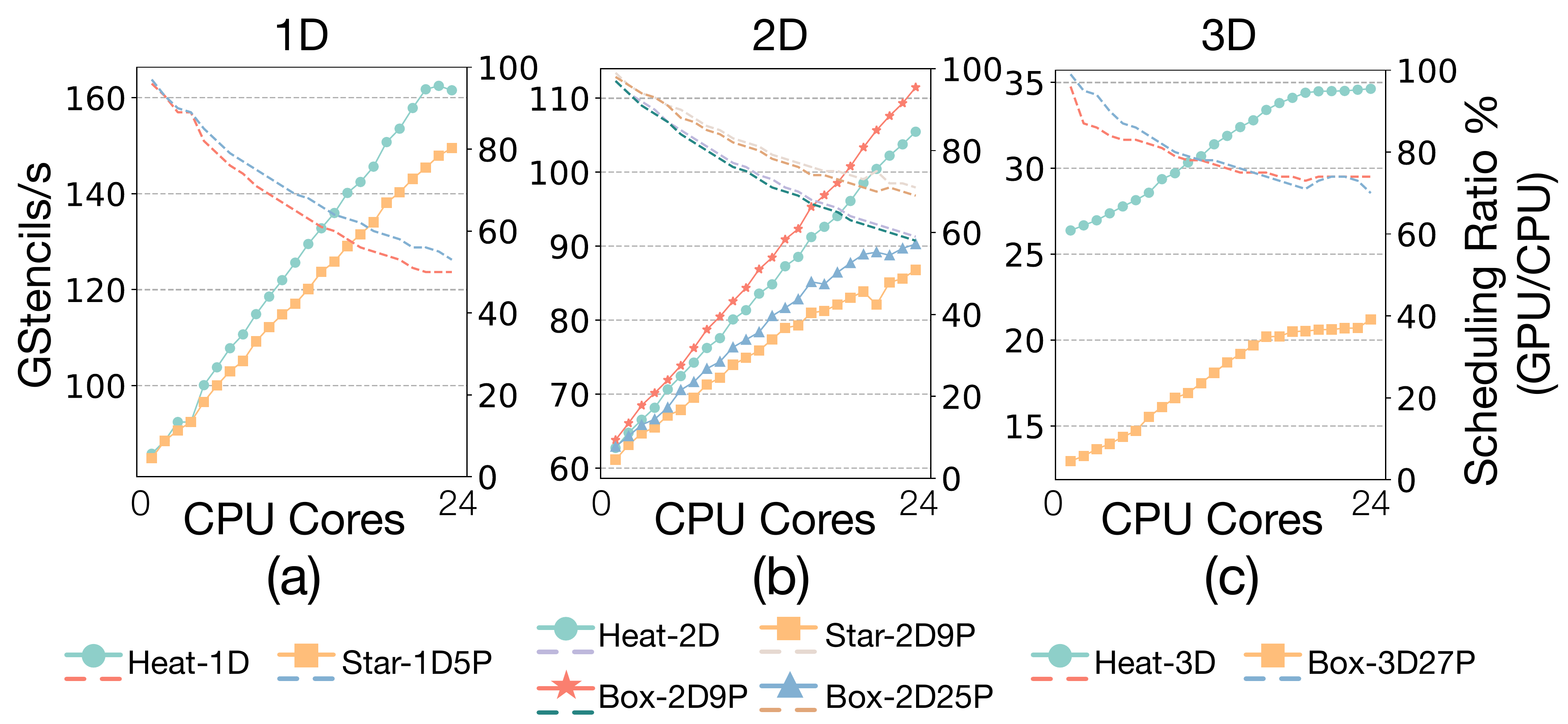}
  \caption{\label{scaling}{Scalability for Tetris with various benchmarks. Solid lines with markers represent the performance, and scheduling ratios (GPU to CPU) of task partition are illustrated with dotted lines.}}
  \end{center} 
  \vspace{-0.5cm}
\end{figure}   

We also evaluate the scalability of Tetris on Cloud for a comprehensive analysis. The detailed parameters are given in Table~\ref{parameters}, where all problem sizes are large enough to approach the scale in a real simulation.  

It can be observed from Figure~\ref{scaling} that Tetris could achieve a good scaling with increased CPU cores. In 1D and 2D benchmarks, nearly linear scalings on Cloud are obtained by Tetris and the CPU with 24 cores provides a competitive performance to GPU in the same order of magnitude.
Surprisingly, a 49.9\% scheduling ration is determined by Tetris, which demonstrates a scalable scheduler design and breaks the stereotype of CPU inferior to GPU.
With the increase of the benchmark dimension, the scalability drops slightly when more CPU cores are employed due to the inherent complexity for multidimensional computations. Moreover, it has to perform a great deal of coordination among CPU cores, incurring communication overheads inevitably.

\subsection{Case Study: Thermal Diffusion  }
\begin{figure}[hbt]
  \begin{center}
  \centering
  \includegraphics[width=0.4\textwidth]{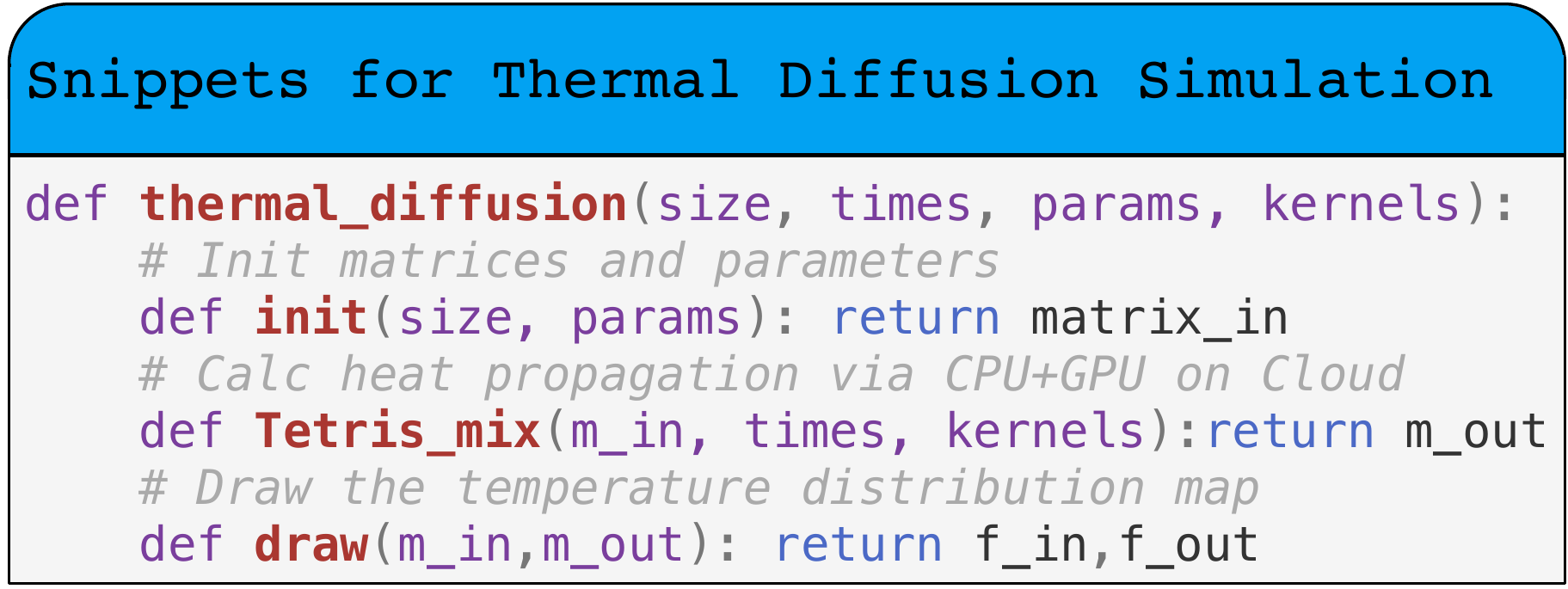}
  \caption{\label{interface}{Code snippets for thermal diffusion simulation with Tetris.}}
  \end{center} 
  \vspace{-0.2in}
\end{figure}

\begin{figure}[hbt]
  \begin{center}
  \centering
  \includegraphics[width=0.45\textwidth]{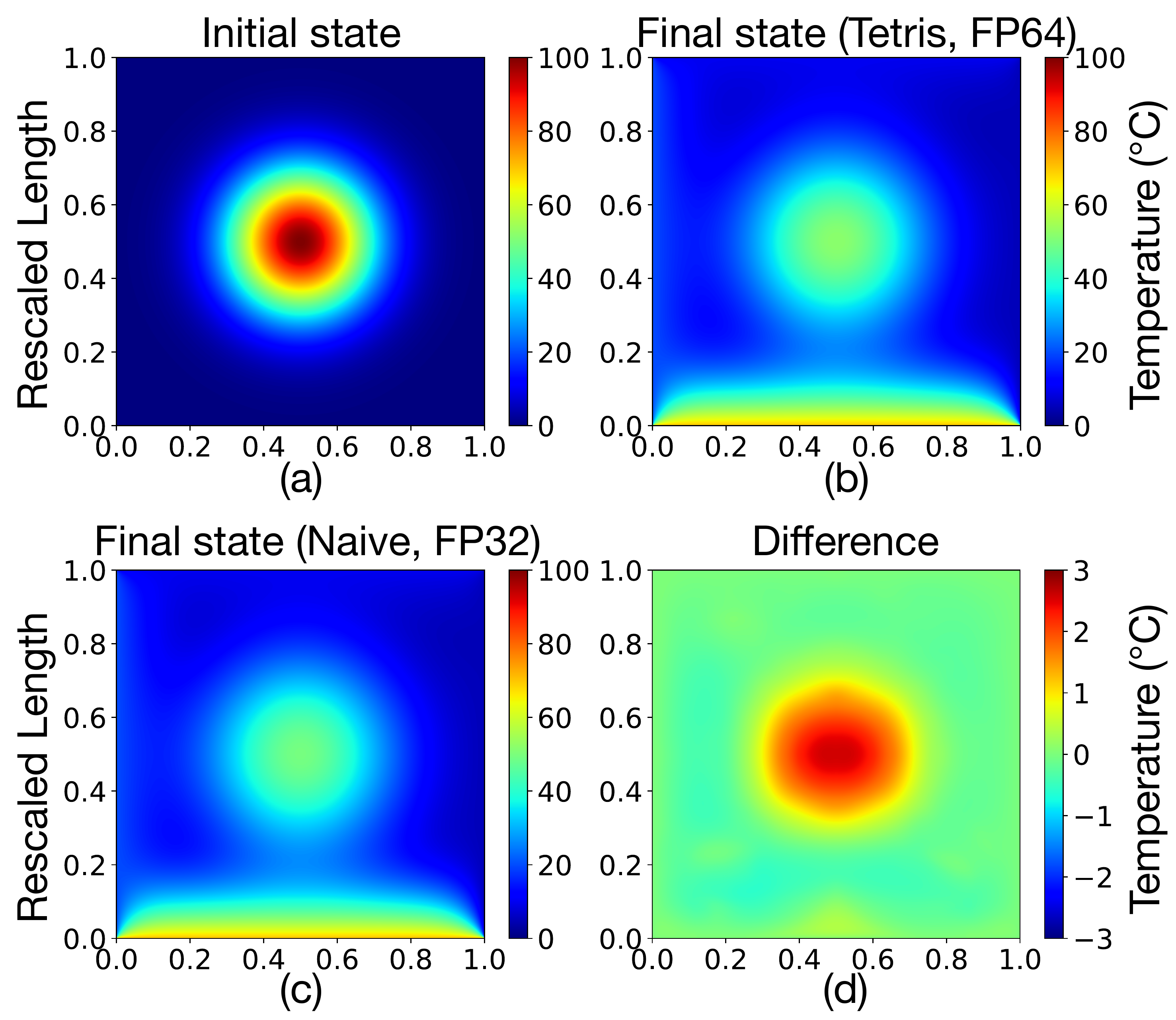}
  \caption{\label{heat_diff}{Visualization of temperature distribution before and
after the thermal diffusion on the square Cu plate. The difference is measured by  absolute errors (the reds: positive errors (hotter), the greens: zero errors, the blues: negative errors (colder)).}}
  \end{center} 
  \vspace{-0.5cm}
\end{figure}

\begin{table}[hbt] \small  
  \caption{Performance improvements by Tetris}\label{label:heat}
  \renewcommand\tabcolsep{2pt} 
  \centering\begin{tabular}{lccr}
  \toprule
  Methods &     Time(s) & Performance (GStencil/s) & Speedup\\ \hline
  Naive &  124,448.5 & 2.8 & -\\
  Tetris (CPU) &   27,709.1 & 14.8 &4.5x\\
  Tetris (GPU) &    5,597.4 & 63.3 &22.6x\\
  Tetris &    4,270.9 & 82.9 &29.6x\\
  \bottomrule
  \end{tabular}
\end{table}

\begin{table}[hbt]\small 
\begin{threeparttable}
\caption{Analytical accuracy comparison$^1$}\label{label:error}
  \renewcommand\tabcolsep{1.5pt} 
\begin{tabular}{lcccccr}
\toprule
 \multirow{2}{*}{Deviation}& \multicolumn{3}{l}{ Absolute Error ($^{\circ}$C)} & \multicolumn{3}{l}{Relative Error (\%)} \\
 &  $>$0.1 & $>$0.5 & $>$1.0 &      $>$1.0 & $>$3.0 & $>$5.0  \\\cmidrule(r){0-0} \cmidrule(r){2-4}  \cmidrule(r){5-7}
Tetris by FP64 (\%)  &    0.0 & 0.0 & 0.0    &     0.0 & 0.0 & 0.0   \\
Naive by FP32 (\%)&   73.1 & 17.2 & 11.3  &     62.1 & 14.3 & 0.2\\
 \bottomrule  
\end{tabular}
\begin{tablenotes}\footnotesize
  \item[1] The accuracy errors are all compared with the Naive method(FP64).  
  \end{tablenotes}
\end{threeparttable}
\vspace{-0.1in}
\end{table}

To further prove the validity of Tetris in real application, simulations of thermal diffusion on a square copper plate (CFL number: $\mu = 0.23$) are performed at 100$^{\circ}$C with dimensions of 15 mm on a side (Grid size: $9,600\times 9600$). The simulation is to calculate the temperature distribution $D$  depending on time for a long-time evolution ($3.8\times 10^6$ time steps), i.e. $D = U(t, x, y)$ simulated by Laplacian numerical calculation using 5-point Stencil finite difference methods in Section~\ref{pde_derivation}. 


Provided with a simple interface illustrated in Figure~\ref{interface}, scientists could democratize large-scale thermal simulation on Cloud with minimal efforts by Tetris. Table~\ref{label:heat} shows the detailed performance improvements on
thermal diffusion by Tetris. The speedup ranges from 4.5x to 29.6x with different versions of Tetris, showing that it could provide a significant benefit over real {scientific computing} applications.

As shown in Figure~\ref{heat_diff}(a), at the beginning of the simulation, it is clear that the initial temperature is a Gaussian, and it will be cooler at the edges with the hottest temperature at the plate center. After a long-time thermal evolution by Tetris, the temperature distribution in Figure~\ref{heat_diff}(b) is radically different from the one obtained at the beginning, and the temperature of plate center decreases to 52.0$^{\circ}$C, which is coherent with the thermodynamic theory~\cite{bluman1969general}. 

It is worth noting that we also perform a comparison experiment with FP32 on the same initial state of the system. The results by FP32 and the differences with Tetris (FP64) are visualized in Figure~\ref{heat_diff}(c) and Figure~\ref{heat_diff}(d) respectively.
Table~\ref{label:error} also lists an analytical accuracy comparison of FP64 and FP32 intuitively. Significantly, an overall 73.1\% deviation is measured for temperatures fluctuated by 0.1$^{\circ}$C, which demonstrates an unignorable variation caused by insufficient accuracies (FP32). 

\section{Related Work}

Known as one of the seven Dwarfs, Stencil is extensively involved in scientific computing~\cite{li2021reducing,asanovic2006landscape}. The representative work can largely be classified in two directions in terms of architectures.

It is acknowledged that {scientific computing} is sensitive to the data precision, and thus a large quantity of work addresses this issue by utilizing FP64 operations on CPU architectures. Traditional approaches have mainly focused on either vectorization or tiling schemes, aiming at improving the in-core data parallelism and the data locality in cache respectively~\cite{yuan2019tessellating}. 
For SIMD implementation on CPU, data alignment conflicts incurred by vectorization is a crucial performance-limiting factor~\cite{li2022efficient,henretty2011data,zhao2019exploiting}. 
Bandishti et al. enhance the Pluto compiler to incorporate a strategy for diamond tiling, and is particularly effective in parallelizing stencil computations~\cite{bandishti2012tiling}.
Folding reduces the data reorganization overhead for vectorization and shows a better performance than Pluto~\cite{bandishti2012tiling}, Tessellation~\cite{yuan2019tessellating}, YASK~\cite{yount2015vector,yount2016yask} and DLT~\cite{henretty2013compiler}, which
is considered as one of state-of-the-arts approach for vectorization. However, the frequent in-CPU transpose easily incurs register spilling~\cite{li2021reducing}.  
Tiling~\cite{Irigoin.Triolet:popl88,McKeller.Coffman:cacm69,Lam+:asplos91,Wolf.Lam:pldi91,Wolfe:sc89} is one of the most powerful transformation techniques to explore the data locality of multiple loop nests. Notable work for Stencil includes hyper-rectangle tiling~\cite{Ding.He:sc01,Rastello.Dauxois:ipdps02,Rivera.Tseng:sc00,Nguyen+:sc10},
time skewed tiling~\cite{Song.Li:pldi99,Wonnacott:ijpp02,Jin+:sc01},
diamond tiling~\cite{bondhugula2008practical,bandishti2012tiling},
cache oblivious Tiling~\cite{10.1145/1989493.1989508,strzodka2010cache,Frigo.Strumpen:ics05}, split-tiling~\cite{henretty2013compiler} and Tessellation~\cite{yuan2019tessellating}, which are mostly compiler transformation techniques. A variety of auto-tuning frameworks \cite{christen2011patus,gysi2015modesto,kamil2010auto,zhang2012auto} have also been presented by using varied hyper-rectangular tiles to exploit data reuse alone. However, redundant computations are involved in these work to resolve the introduced inter-tile dependencies especially when combined with vectorization, which hinder the concurrent execution on different cores.  

With the interests in more efficient solutions, Stencil on GPU is also being exploited in the community~\cite{holewinski2012high,zhang2012auto,grosser2013promises}. Common stencil optimizations on GPU include tiling and loop unrolling~\cite{holewinski2012high,krishnamoorthy2007effective,meng2009performance}, however, correct and efficient implementation of these techniques are challenging because careful utilization of limited registers and shared memory is required for complex data dependencies~\cite{matsumura2020an5d}. Since GPU runs low-precision operations (FP16/FP32) highly faster than high precision (FP64), most existing work employ low-precision designs while they are only applied to a limited set of benchmarks and hardly meet the strict tastes of {scientific computing}~\cite{maruyama2014optimizing,nguyen20103}.    
Zhao presents a general stencil framework by exploiting data reuse within a block~\cite{zhao2019exploiting}, however, the algorithms are implemented separately on CPUs and GPUs in practice. AN5D achieves high-degree temporal blocking on GPUs, while it only conducts the parameter search with single-precision, and the method is limited to CUDA cores~\cite{matsumura2020an5d}. 


\section{Conclusion}
This paper presents Tetris, a disruptive system that democratizes {Stencil-driven scientific computing} on CPU and GPU of Cloud with polymorphic tiling tetrominoes to tessellate the spatial and temporal dimensions perfectly. Sophisticated Pattern Mapping by register-level tetrominoes, efficient Locality Enhancer by cache/SMEM-level tetrominoes, and novel concurrent scheduler by memory-level tetrominoes are also first proposed to improve the performance on heterogeneous architecture. Our promising results demonstrate that Tetris can achieve an unprecedented performance improvement compared to the existing state-of-the-arts.

\bibliographystyle{acm}
\bibliography{ref.bib}

\begin{thebibliography}{10}

\bibitem{ahmad2021fast}
{\sc Ahmad, Z., Chowdhury, R., Das, R., Ganapathi, P., Gregory, A., and Zhu,
  Y.}
\newblock Fast stencil computations using fast fourier transforms.
\newblock In {\em Proceedings of the 33rd ACM Symposium on Parallelism in
  Algorithms and Architectures\/} (2021), pp.~8--21.

\bibitem{ando2021digital}
{\sc Ando, K., Bale, R., Li, C., Matsuoka, S., Onishi, K., and Tsubokura, M.}
\newblock Digital transformation of droplet/aerosol infection risk assessment
  realized on" fugaku" for the fight against covid-19.
\newblock {\em arXiv preprint arXiv:2110.09769\/} (2021).

\bibitem{asanovic2006landscape}
{\sc Asanovic, K., Bodik, R., Catanzaro, B.~C., Gebis, J.~J., Husbands, P.,
  Keutzer, K., Patterson, D.~A., Plishker, W.~L., Shalf, J., Williams, S.~W.,
  et~al.}
\newblock The landscape of parallel computing research: A view from berkeley.

\bibitem{asanovic2008parallel}
{\sc Asanovic, K., Bodik, R., Demmel, J., Keaveny, T., Keutzer, K.,
  Kubiatowicz, J.~D., Lee, E.~A., Morgan, N., Necula, G., Patterson, D.~A.,
  et~al.}
\newblock The parallel computing laboratory at uc berkeley: A research agenda
  based on the berkeley view.
\newblock {\em EECS Department, University of California, Berkeley, Tech.
  Rep\/} (2008).

\bibitem{bailey2005high}
{\sc Bailey, D.~H.}
\newblock High-precision floating-point arithmetic in scientific computation.
\newblock {\em Computing in science \& engineering 7}, 3 (2005), 54--61.

\bibitem{bailey2012high}
{\sc Bailey, D.~H., Barrio, R., and Borwein, J.~M.}
\newblock High-precision computation: Mathematical physics and dynamics.
\newblock {\em Applied Mathematics and Computation 218}, 20 (2012),
  10106--10121.

\bibitem{bandishti2012tiling}
{\sc Bandishti, V., Pananilath, I., and Bondhugula, U.}
\newblock Tiling stencil computations to maximize parallelism.
\newblock In {\em SC'12: Proceedings of the International Conference on High
  Performance Computing, Networking, Storage and Analysis\/} (2012), IEEE,
  pp.~1--11.

\bibitem{bluman1969general}
{\sc Bluman, G.~W., and Cole, J.~D.}
\newblock The general similarity solution of the heat equation.
\newblock {\em Journal of mathematics and mechanics 18}, 11 (1969), 1025--1042.

\bibitem{bondhugula2008practical}
{\sc Bondhugula, U., Hartono, A., Ramanujam, J., and Sadayappan, P.}
\newblock A practical automatic polyhedral parallelizer and locality optimizer.
\newblock In {\em Proceedings of the 29th ACM SIGPLAN Conference on Programming
  Language Design and Implementation\/} (2008), pp.~101--113.

\bibitem{cai2021physics}
{\sc Cai, S., Wang, Z., Wang, S., Perdikaris, P., and Karniadakis, G.~E.}
\newblock Physics-informed neural networks for heat transfer problems.
\newblock {\em Journal of Heat Transfer 143}, 6 (2021).

\bibitem{christen2011patus}
{\sc Christen, M., Schenk, O., and Burkhart, H.}
\newblock Patus: A code generation and autotuning framework for parallel
  iterative stencil computations on modern microarchitectures.
\newblock In {\em IPDPS 2011}, IEEE.

\bibitem{christen2012patus}
{\sc Christen, M., Schenk, O., and Cui, Y.}
\newblock Patus for convenient high-performance stencils: evaluation in
  earthquake simulations.
\newblock In {\em SC'12: Proceedings of the International Conference on High
  Performance Computing, Networking, Storage and Analysis\/} (2012), IEEE,
  pp.~1--10.

\bibitem{dakkak2019accelerating}
{\sc Dakkak, A., Li, C., Xiong, J., Gelado, I., and Hwu, W.-m.}
\newblock Accelerating reduction and scan using tensor core units.
\newblock In {\em Proceedings of the ACM International Conference on
  Supercomputing\/} (2019), pp.~46--57.

\bibitem{datta2008stencil}
{\sc Datta, K., Murphy, M., Volkov, V., Williams, S., Carter, J., Oliker, L.,
  Patterson, D., Shalf, J., and Yelick, K.}
\newblock Stencil computation optimization and auto-tuning on state-of-the-art
  multicore architectures.
\newblock In {\em SC'08: Proceedings of the 2008 ACM/IEEE conference on
  Supercomputing\/} (2008), IEEE, pp.~1--12.

\bibitem{denzler2021casper}
{\sc Denzler, A., Bera, R., Hajinazar, N., Singh, G., Oliveira, G.~F.,
  G{\'o}mez-Luna, J., and Mutlu, O.}
\newblock Casper: Accelerating stencil computation using near-cache processing.
\newblock {\em arXiv preprint arXiv:2112.14216\/} (2021).

\bibitem{Ding.He:sc01}
{\sc Ding, C., and He, Y.}
\newblock A ghost cell expansion method for reducing communications in solving
  pde problems.
\newblock SC '01, pp.~50--50.

\bibitem{Frigo.Strumpen:ics05}
{\sc Frigo, M., and Strumpen, V.}
\newblock Cache oblivious stencil computations.
\newblock ICS '05, pp.~361--366.

\bibitem{fu20179}
{\sc Fu, H., He, C., Chen, B., Yin, Z., Zhang, Z., Zhang, W., Zhang, T., Xue,
  W., Liu, W., Yin, W., et~al.}
\newblock 18.9-pflops nonlinear earthquake simulation on sunway taihulight:
  enabling depiction of 18-hz and 8-meter scenarios.
\newblock In {\em Proceedings of the International Conference for High
  Performance Computing, Networking, Storage and Analysis\/} (2017), pp.~1--12.

\bibitem{fu2017redesigning}
{\sc Fu, H., Liao, J., Ding, N., Duan, X., Gan, L., Liang, Y., Wang, X., Yang,
  J., Zheng, Y., Liu, W., et~al.}
\newblock Redesigning cam-se for peta-scale climate modeling performance and
  ultra-high resolution on sunway taihulight.
\newblock In {\em Proceedings of the International Conference for High
  Performance Computing, Networking, Storage and Analysis\/} (2017), pp.~1--12.

\bibitem{granlund2012instruction}
{\sc Granlund, T.}
\newblock Instruction latencies and throughput for amd and intel x86
  processors.
\newblock In {\em Technical report}. KTH, 2012.

\bibitem{grosser2013promises}
{\sc Grosser, T., Verdoolaege, S., Cohen, A., and Sadayappan, P.}
\newblock {\em The Promises of Hybrid Hexagonal/Classical Tiling for GPU}.
\newblock PhD thesis, INRIA, 2013.

\bibitem{guide21url}
{\sc Guide, I.~I.}
\newblock Url: https://software. intel. com/sites/landingpage.
\newblock {\em IntrinsicsGuide (access date: 21.10. 2019) 78\/}.

\bibitem{gysi2015modesto}
{\sc Gysi, T., Grosser, T., and Hoefler, T.}
\newblock Modesto: Data-centric analytic optimization of complex stencil
  programs on heterogeneous architectures.
\newblock In {\em ICS 2015\/} (2015), pp.~177--186.

\bibitem{henretty2011data}
{\sc Henretty, T., Stock, K., Pouchet, L.-N., Franchetti, F., Ramanujam, J.,
  and Sadayappan, P.}
\newblock Data layout transformation for stencil computations on short-vector
  simd architectures.
\newblock In {\em International Conference on Compiler Construction\/} (2011),
  Springer, pp.~225--245.

\bibitem{henretty2013compiler}
{\sc Henretty, T., Veras, R., Franchetti, F., Pouchet, L.-N., Ramanujam, J.,
  and Sadayappan, P.}
\newblock A stencil compiler for short-vector simd architectures.
\newblock In {\em Proceedings of the 27th International ACM Conference on
  International Conference on Supercomputing\/} (New York, NY, USA, 2013), ICS
  '13, Association for Computing Machinery, pp.~13--24.

\bibitem{holewinski2012high}
{\sc Holewinski, J., Pouchet, L.-N., and Sadayappan, P.}
\newblock High-performance code generation for stencil computations on gpu
  architectures.
\newblock In {\em Proceedings of the 26th ACM international conference on
  Supercomputing\/} (2012), pp.~311--320.

\bibitem{Irigoin.Triolet:popl88}
{\sc Irigoin, F., and Triolet, R.}
\newblock Supernode partitioning.
\newblock POPL '88, pp.~319--329.

\bibitem{jia2020pushing}
{\sc Jia, W., Wang, H., Chen, M., Lu, D., Lin, L., Car, R., Weinan, E., and
  Zhang, L.}
\newblock Pushing the limit of molecular dynamics with ab initio accuracy to
  100 million atoms with machine learning.
\newblock In {\em SC20: International conference for high performance
  computing, networking, storage and analysis\/} (2020), IEEE, pp.~1--14.

\bibitem{Jin+:sc01}
{\sc Jin, G., Mellor-Crummey, J., and Fowler, R.}
\newblock Increasing temporal locality with skewing and recursive blocking.
\newblock SC '01, pp.~43--43.

\bibitem{kamil2010auto}
{\sc Kamil, S., Chan, C., Oliker, L., Shalf, J., and Williams, S.}
\newblock An auto-tuning framework for parallel multicore stencil computations.
\newblock In {\em 2010 IEEE International Symposium on Parallel \& Distributed
  Processing (IPDPS)\/} (2010), IEEE, pp.~1--12.

\bibitem{krishnamoorthy2007effective}
{\sc Krishnamoorthy, S., Baskaran, M., Bondhugula, U., Ramanujam, J., Rountev,
  A., and Sadayappan, P.}
\newblock Effective automatic parallelization of stencil computations.
\newblock {\em ACM sigplan notices 42}, 6 (2007), 235--244.

\bibitem{Lam+:asplos91}
{\sc Lam, M.~D., Rothberg, E.~E., and Wolf, M.~E.}
\newblock The cache performance and optimizations of blocked algorithms.
\newblock ASPLOS IV, pp.~63--74.

\bibitem{li2019openkmc}
{\sc Li, K., Shang, H., Zhang, Y., Li, S., Wu, B., Wang, D., Zhang, L., Li, F.,
  Chen, D., and Wei, Z.}
\newblock Openkmc: a kmc design for hundred-billion-atom simulation using
  millions of cores on sunway taihulight.
\newblock In {\em Proceedings of the International Conference for High
  Performance Computing, Networking, Storage and Analysis\/} (2019), pp.~1--16.

\bibitem{li2021reducing}
{\sc Li, K., Yuan, L., Zhang, Y., and Yue, Y.}
\newblock Reducing redundancy in data organization and arithmetic calculation
  for stencil computations.
\newblock In {\em Proceedings of the International Conference for High
  Performance Computing, Networking, Storage and Analysis\/} (2021), pp.~1--15.

\bibitem{li2022efficient}
{\sc Li, K., Yuan, L., Zhang, Y., Yue, Y., and Cao, H.}
\newblock An efficient vectorization scheme for stencil computation.
\newblock In {\em 2022 IEEE International Parallel and Distributed Processing
  Symposium (IPDPS)\/} (2022), IEEE, pp.~650--660.

\bibitem{maruyama2014optimizing}
{\sc Maruyama, N., and Aoki, T.}
\newblock Optimizing stencil computations for nvidia kepler gpus.
\newblock In {\em Proceedings of the 1st international workshop on
  high-performance stencil computations, Vienna\/} (2014), Citeseer,
  pp.~89--95.

\bibitem{matsumura2020an5d}
{\sc Matsumura, K., Zohouri, H.~R., Wahib, M., Endo, T., and Matsuoka, S.}
\newblock An5d: automated stencil framework for high-degree temporal blocking
  on gpus.
\newblock In {\em Proceedings of the 18th ACM/IEEE International Symposium on
  Code Generation and Optimization\/} (2020), pp.~199--211.

\bibitem{McKeller.Coffman:cacm69}
{\sc McKellar, A.~C., and Coffman, Jr., E.~G.}
\newblock Organizing matrices and matrix operations for paged memory systems.
\newblock {\em Commun. ACM 12}, 3 (1969), 153--165.

\bibitem{meng2009performance}
{\sc Meng, J., and Skadron, K.}
\newblock Performance modeling and automatic ghost zone optimization for
  iterative stencil loops on gpus.
\newblock In {\em Proceedings of the 23rd international conference on
  Supercomputing\/} (2009), pp.~256--265.

\bibitem{heatformula}
{\sc MIT}.
\newblock Stencil computing, 2010.
\newblock [Online; accessed 29-Nov-2022].

\bibitem{Nguyen+:sc10}
{\sc Nguyen, A., Satish, N., Chhugani, J., Kim, C., and Dubey, P.}
\newblock 3.5-d blocking optimization for stencil computations on modern cpus
  and gpus.
\newblock SC '10, pp.~1--13.

\bibitem{nguyen20103}
{\sc Nguyen, A., Satish, N., Chhugani, J., Kim, C., and Dubey, P.}
\newblock 3.5-d blocking optimization for stencil computations on modern cpus
  and gpus.
\newblock In {\em SC'10: Proceedings of the 2010 ACM/IEEE International
  Conference for High Performance Computing, Networking, Storage and
  Analysis\/} (2010), IEEE, pp.~1--13.

\bibitem{CUDAToolkit}
{\sc Nvidia}.
\newblock Cuda toolkit v11.8.0, 2022.
\newblock [Online; accessed 29-Nov-2022].

\bibitem{grandchallenge}
{\sc Nvidia}.
\newblock Solve the world’s greatest challenges with supercomputing, 2022.
\newblock [Online; accessed 29-Nov-2022].

\bibitem{ogbole2021cloud}
{\sc Ogbole, M.~O., Ogbole, E.~A., and Olagesin, A.}
\newblock Cloud systems and applications: A review.
\newblock {\em International Journal of Scientific Research in Computer
  Science, Engineering and Information Technology\/} (2021), 142--149.

\bibitem{GrandChallenges}
{\sc on~Physical Mathematical Engineering~Science, T.~C.}
\newblock Grand challenges: High performance computing and communications.
\newblock Tech. rep., The FY 1992 US Research and Development Program, 1992.

\bibitem{pisha2021accelerating}
{\sc Pisha, L., and Ligowski, {\L}.}
\newblock Accelerating non-power-of-2 size fourier transforms with gpu tensor
  cores.
\newblock In {\em 2021 IEEE International Parallel and Distributed Processing
  Symposium (IPDPS)\/} (2021), IEEE, pp.~507--516.

\bibitem{raissi2019physics}
{\sc Raissi, M., Perdikaris, P., and Karniadakis, G.~E.}
\newblock Physics-informed neural networks: A deep learning framework for
  solving forward and inverse problems involving nonlinear partial differential
  equations.
\newblock {\em Journal of Computational physics 378\/} (2019), 686--707.

\bibitem{Rastello.Dauxois:ipdps02}
{\sc Rastello, F., and Dauxois, T.}
\newblock Efficient tiling for an ode discrete integration program: Redundant
  tasks instead of trapezoidal shaped-tiles.
\newblock IPDPS '02, pp.~138--.

\bibitem{ren2021zero}
{\sc Ren, J., Rajbhandari, S., Aminabadi, R.~Y., Ruwase, O., Yang, S., Zhang,
  M., Li, D., and He, Y.}
\newblock $\{$ZeRO-Offload$\}$: Democratizing $\{$Billion-Scale$\}$ model
  training.
\newblock In {\em 2021 USENIX Annual Technical Conference (USENIX ATC 21)\/}
  (2021), pp.~551--564.

\bibitem{Rivera.Tseng:sc00}
{\sc Rivera, G., and Tseng, C.-W.}
\newblock Tiling optimizations for 3d scientific computations.
\newblock SC '00.

\bibitem{soh2020microsoft}
{\sc Soh, J., Copeland, M., Puca, A., and Harris, M.}
\newblock Microsoft azure and cloud computing.
\newblock In {\em Microsoft Azure}. Springer, 2020, pp.~3--20.

\bibitem{Song.Li:pldi99}
{\sc Song, Y., and Li, Z.}
\newblock New tiling techniques to improve cache temporal locality.
\newblock PLDI '99, pp.~215--228.

\bibitem{strzodka2010cache}
{\sc Strzodka, R., Shaheen, M., Pajak, D., and Seidel, H.-P.}
\newblock Cache oblivious parallelograms in iterative stencil computations.
\newblock In {\em Proceedings of the 24th ACM International Conference on
  Supercomputing\/} (2010), pp.~49--59.

\bibitem{10.1145/1989493.1989508}
{\sc Tang, Y., Chowdhury, R.~A., Kuszmaul, B.~C., Luk, C.-K., and Leiserson,
  C.~E.}
\newblock The pochoir stencil compiler.
\newblock In {\em Proceedings of the Twenty-Third Annual ACM Symposium on
  Parallelism in Algorithms and Architectures\/} (New York, NY, USA, 2011),
  SPAA '11, Association for Computing Machinery, pp.~117--128.

\bibitem{Wolf.Lam:pldi91}
{\sc Wolf, M.~E., and Lam, M.~S.}
\newblock A data locality optimizing algorithm.
\newblock PLDI '91, pp.~30--44.

\bibitem{Wolfe:sc89}
{\sc Wolfe, M.}
\newblock More iteration space tiling.
\newblock Supercomputing '89, pp.~655--664.

\bibitem{Wonnacott:ijpp02}
{\sc Wonnacott, D.}
\newblock Achieving scalable locality with time skewing.
\newblock {\em Int. J. Parallel Program. 30}, 3 (June 2002), 181--221.

\bibitem{xiao2018communication}
{\sc Xiao, J., Li, S., Wu, B., Zhang, H., Li, K., Yao, E., Zhang, Y., and Tan,
  G.}
\newblock Communication-avoiding for dynamical core of atmospheric general
  circulation model.
\newblock In {\em Proceedings of the 47th International Conference on Parallel
  Processing\/} (2018), pp.~1--10.

\bibitem{xushun2021}
{\sc Xu, S., Wang, W., Zhang, J., Jiang, J., Jin, Z., and Chi, X.}
\newblock High performance computing algorithms and software for heterogeneous
  computing.
\newblock {\em Journal of Software 32}, 8 (2021), 2365--2376.

\bibitem{yount2015vector}
{\sc Yount, C.}
\newblock Vector folding: Improving stencil performance via multi-dimensional
  simd-vector representation.
\newblock In {\em 2015 IEEE 17th International Conference on High Performance
  Computing and Communications, 2015 IEEE 7th International Symposium on
  Cyberspace Safety and Security, and 2015 IEEE 12th International Conference
  on Embedded Software and Systems\/} (2015), IEEE, pp.~865--870.

\bibitem{yount2016yask}
{\sc Yount, C., Tobin, J., Breuer, A., and Duran, A.}
\newblock Yask-yet another stencil kernel: A framework for hpc stencil
  code-generation and tuning.
\newblock In {\em 2016 Sixth International Workshop on Domain-Specific
  Languages and High-Level Frameworks for High Performance Computing
  (WOLFHPC)\/} (2016), IEEE, pp.~30--39.

\bibitem{yuan2019tessellating}
{\sc Yuan, L., Huang, S., Zhang, Y., and Cao, H.}
\newblock Tessellating star stencils.
\newblock In {\em Proceedings of the 48th International Conference on Parallel
  Processing\/} (2019), pp.~1--10.

\bibitem{10.1145/3126908.3126920}
{\sc Yuan, L., Zhang, Y., Guo, P., and Huang, S.}
\newblock Tessellating stencils.
\newblock In {\em Proceedings of the International Conference for High
  Performance Computing, Networking, Storage and Analysis\/} (New York, NY,
  USA, 2017), SC '17, Association for Computing Machinery.

\bibitem{zhang2012auto}
{\sc Zhang, Y., and Mueller, F.}
\newblock Auto-generation and auto-tuning of 3d stencil codes on gpu clusters.
\newblock In {\em Proceedings of the Tenth International Symposium on Code
  Generation and Optimization\/} (2012), pp.~155--164.

\bibitem{zhao2019exploiting}
{\sc Zhao, T., Basu, P., Williams, S., Hall, M., and Johansen, H.}
\newblock Exploiting reuse and vectorization in blocked stencil computations on
  cpus and gpus.
\newblock In {\em SC'19\/} (2019), pp.~1--44.

\end{thebibliography}


\end{document}